\documentclass{lmcs}
\pdfoutput=1

\usepackage{lastpage}
\lmcsdoi{21}{1}{5}
\lmcsheading{}{\pageref{LastPage}}{}{}%
{Sep.~01,~2022}{Jan.~22,~2025}{}

\usepackage[utf8]{inputenc}
\usepackage{marvosym}

\usepackage{stmaryrd}
\usepackage{amssymb} 
\usepackage{array}
\usepackage{listings}
\usepackage{ebproof}
\usepackage{amsmath}
\usepackage{subcaption}
\usepackage{tikz}
\usetikzlibrary{matrix}
\usetikzlibrary{decorations.pathreplacing,calc}
\newcommand{\tikzmark}[1]{\tikz[overlay,remember picture] \node (#1) {};}
\usepackage{booktabs}

\newcommand{\kwfont}[1]{\texttt{\bfseries{\upshape{#1}}}}
\newcommand{\ens}[1]{\mathbb{#1}}
\newcommand{\expr}[1]{\mathtt{#1}}
\newcommand{\e}{\expr{e}}
\newcommand{\Operator}{\ens{O}}
\newcommand{\Var}{\ens{V}}
\newcommand{\oper}[1]{\mathtt{#1}}
\newcommand{\op}{\oper{op}}
\newcommand{\variable}[1]{\mathtt{#1}}
\let\vv\relax
\newcommand{\vv}{\variable{v}}
\newcommand{\uu}{\variable{u}}

\newcommand{\x}{\variable{x}}
\newcommand{\y}{\variable{y}}
\newcommand{\z}{\variable{z}}
\newcommand{\va}{\variable{a}}
\newcommand{\vb}{\variable{b}}
\newcommand{\term}{\variable{t}}
\newcommand{\pvterm}{\variable{t}}
\newcommand{\clos}{\variable{c}}
\newcommand{\X}{\variable{X}}
\newcommand{\Y}{\variable{Y}}

\newcommand{\Command}[1]{\mathtt{ #1}}
\newcommand{\cmd}{\Command{st}}
\newcommand{\stt}{\Command{body}}
\newcommand{\param}{\Command{param}}
\newcommand{\procname}{\mathtt{P}}
\newcommand{\callcc}{\kwfont{call}}
\newcommand{\prog}{\mathtt{prg}}
\newcommand{\boite}[2]{\kwfont{box}\ \kwfont{[}#1\kwfont{]}\ \kwfont{in}\ #2}

\newcommand{\procedure}{\mathtt{p}}
\newcommand{\instr}[1]{\mathtt{#1}}
\newcommand{\skp}{\kwfont{skip}}
\newcommand{\sep}{\instr{;\ }}
\newcommand{\asg}{\instr{\ :=\ }}
\newcommand{\while}{\kwfont{while}}
\newcommand{\ret}{\kwfont{return\ }}
\newcommand{\ifa}{\kwfont{if}}

\newcommand{\elsea}{\kwfont{\ else\ }}
\newcommand{\rgl}{::=}
\newcommand{\FV}{\Var}

\newcommand{\true}{1}
\newcommand{\false}{0}
\newcommand{\W}{\mathcal{W}}
\newcommand{\pvW}{\mathcal{W}}
\newcommand{\TW}{{\tt W}}
\newcommand{\STW}{{\tt T}}
\newcommand{\SN}{\mathrm{SN}}
\newcommand{\SCP}{\mathrm{SCP}}
\newcommand{\SCPS}{\mathrm{SCP}_{\mathrm{S}}}
\newcommand{\SST}{\mathcal{T}_\TW}
\newcommand{\w}{\mathit{w}}
\newcommand{\vi}{\mathit{v}}
\newcommand{\ui}{\mathit{u}}
\newcommand{\size}[1]{|#1|}
\newcommand{\sem}[1]{\mbox{$\llbracket #1 \rrbracket$}}
\newcommand{\store}{\mu}

\newcommand{\Imp}{,}
\newcommand{\tier}[1]{\mathbf{#1}}
\newcommand{\letin}[1]{\kwfont{declare}\ #1\ \kwfont{in}\ }
\newcommand{\tiera}{\tier{0}}
\newcommand{\tierb}{\tier{1}}
\newcommand{\tierc}{\tier{2}}

\newcommand{\slat}[1]{\mathbf{#1}}
\newcommand{\sla}{\slat{k}}
\newcommand{\slb}{\slat{k'}}
\newcommand{\SL}{\mathbf{N}}
\newcommand{\meet}{\min}
\newcommand{\mini}{\meet}
\newcommand{\join}{\max}
\newcommand{\maxi}{\join}
\newcommand{\ord}{\leq}
\newcommand{\ordst}{<}
\newcommand{\pbl}{\typenv, \typop \vdash}
\newcommand{\pbla}{\typenvs,\typproc,\typop \vdash}
\newcommand{\pblap}{\diamond_{\X,\y}^{\TW^\TW,\TW}\vdash}
\newcommand{\pblapp}{\diamond_{\X,\y,\x}^{\TW^\TW,\overline{\TW}}\vdash}
\newcommand{\dord}{\unlhd}

\newcommand{\typenv}{\Gamma}
\newcommand{\typenvs}{\Gamma_\TW}
\newcommand{\typop}{\Delta}
\newcommand{\typproc}{\Omega}
\newcommand{\dom}{\textit{dom}}
\newcommand{\meq}[1]{{\tt !\!=}_{#1}}
\newcommand{\mequ}[1]{{\tt ==}_{#1}}
\newcommand{\head}{{\tt head}}
\newcommand{\mpred}{\oper{pred}}
\newcommand{\msuc}[1]{\oper{suc}_{#1}}

\newcommand{\mpt}{\mathrm{MPT}}

\newcommand{\safe}{\mathrm{SAFE}}
\newcommand{\st}{\mathrm{ST}}
\newcommand{\FP}{{\tt FP}}
\newcommand{\Ptime}{{\tt P}}
\newcommand{\PSpace}{{\tt PSpace}}
\newcommand{\BFF}{{\tt BFF}}
\newcommand{\toenv}{\to_{\tt env}}
\newcommand{\toexp}{\to_{\tt exp}}
\newcommand{\order}{{\tt ord}}
\newcommand{\tost}{\to_{\tt st}}
\newcommand{\var}{\kwfont{var}\ }
\newcommand{\local}{{\tt local}}

\newcommand{\n}{\mathtt{n}}
\newcommand{\emptyword}{\varepsilon}
\newcommand{\vide}{\epsilon}
\newcommand{\Fun}{\mathbf{F}}

\def\ie{{\em i.e.}}
\def\cf{{\em cf.}}

\newcommand*{\AddNote}[4]{%
    \begin{tikzpicture}[overlay, remember picture]
        \draw [decoration={brace,amplitude=0.5em},decorate,ultra thick,black!60]
            ($(#3)!([yshift=1.5ex]#1)!($(#3)-(0,1)$)$) --  
            ($(#3)!(#2)!($(#3)-(0,1)$)$)
                node [align=left, text width=3cm, pos=0.5, anchor=west] {\quad #4};
    \end{tikzpicture}
}%


\begin{document}
\title[Complete and tractable machine-independent characterizations of $\BFF$]{Complete and tractable machine-independent characterizations of second-order polytime}

\author[E.~Hainry]{Emmanuel Hainry\lmcsorcid{0000-0002-9750-0460}}[a]
\address{Universit{\'e} de Lorraine, CNRS, Inria, LORIA, F-54000 Nancy, France}	
\email{emmanuel.hainry@loria.fr, jean-yves.marion@loria.fr, romain.pechoux@loria.fr}

\author[B.~Kapron]{Bruce M. Kapron\lmcsorcid{0000-0002-3295-543X}}[b]
\address{University of Victoria, Victoria, BC, Canada}
\email{bmkapron@uvic.ca}  

\author[J-Y.~Marion]{Jean-Yves Marion\lmcsorcid{0009-0002-8262-3887}}[a]

\author[R.~P\'echoux]{Romain P\'echoux\lmcsorcid{0000-0003-0601-5425}}[a]


\begin{abstract}
The class of Basic Feasible Functionals $\BFF$ is the second-order counterpart of the class of first-order functions computable in polynomial time.
We present several implicit characterizations of $\BFF$ based on a typed programming language of terms. These terms may perform calls to non-recursive imperative procedures. The type discipline has two layers: the terms follow a standard simply-typed discipline and the procedures follow a standard tier-based type discipline. $\BFF$ consists exactly of the second-order functionals that are computed by typable and terminating programs. The completeness of this characterization surprisingly still holds in the absence of lambda-abstraction and is obtained through the use of closures. Moreover, the termination requirement can be specified as a completeness-preserving instance, which can be decided in time quadratic in the size of the program. As typing is decidable in polynomial time, we obtain the first tractable (\ie, decidable in polynomial time), sound, complete, and implicit characterization of $\BFF$, thus solving a problem open for more than 20 years.
\end{abstract}

\maketitle

\section{Introduction}
\subsection{Motivations}
The class of second-order functions computable in polynomial time was introduced and studied by Mehlhorn~\cite{M76}, building on an earlier proposal by Constable~\cite{Cons73}. Kapron and Cook characterized this class using oracle Turing machines, giving it the name Basic Feasible Functionals ($\BFF$):
\begin{defi}[\cite{KC91}]
Let the size of an oracle $f$ be the first-order function defined by $\size{f}(n) = \max_{\size{y} \leq n}\size{f(y)}$.
A functional F is in $\BFF$, if there are an oracle Turing machine $M$ and a second-order polynomial $P$ such that $M$ computes $F$ in time bounded by $P(\size{f},\size{x})$, for any oracle $f$ and any input $x$.
\end{defi}
The above characterization, that we take as a definition of $\BFF$, shows that functions in $\BFF$ correspond to functionals computed by oracle Turing machines running in polynomial time. The polynomial time bound has a second-order nature since it takes a first-order function, the size of the oracle, as input.  
Since then,  $\BFF$ was by consensus considered as the natural extension to second-order of the well-known class of (first-order) polynomial time computable functions, $\FP$. Notions of second-order polynomial time, while of intrinsic interest, have also been applied in a range of areas, including structural complexity theory~\cite{M76}, resource-bounded topology~\cite{Town90}, complexity of total search problems~\cite{BCEIP98},  feasible real analysis~\cite{KC12}, and verification~\cite{ACG12}.

Starting with Cobham's seminal work~\cite{Cob65}, there have been several attempts to provide  \emph{machine-independent} characterizations of complexity classes  such as ($\Ptime$ and) $\FP$, that is, characterizations based on programming languages rather than on machines. Beyond the purely theoretical aspects, the practical interest of such characterizations is to be able to automatically guarantee that a program can be executed efficiently and in a secure environment. For these characterizations to hold, some restrictions are placed on a given programming language. They ensure that a program can be simulated by a Turing machine in polynomial time and, therefore, corresponds to a function in $\FP$. This property is called \emph{soundness}. Conversely, we would like any function in $\FP$ to be computable by a program satisfying the restrictions. This property is called extensional \emph{completeness}. For automation to be possible, it is necessary that the characterizations studied be \emph{tractable}; that is, decidable in polynomial time. Moreover, they should preferably not require a prior knowledge of the program complexity. One speaks then of \emph{implicit} characterization insofar as the programmer does not have to know an explicit bound on the complexity of the analyzed programs.

In the first-order setting, different restrictions and techniques have been developed to characterize the complexity class $\FP$. One can think, among others, of the safe recursion and ramified recursion techniques for function algebras~\cite{BelCoo92,LeivantMar93}, of interpretation methods for term rewrite systems~\cite{BMM11}, or of light and soft linear logics typing-discipline for lambda-calculi~\cite{Girard98,BT04,BM10}.

In the second-order setting, a machine-independent characterization of $\BFF$ was provided in~\cite{HKMP20}. This characterization uses the tier-based (\ie, safe/ramified recursion-based) type discipline introduced in~\cite{M11} on imperative programs  for characterizing $\FP$ and can be restated as follows: $$ \BFF = \lambda(\sem{\st})_2.$$
$\sem{\st}$ denotes the set of functions computed by typable and terminating programs; $\lambda$ denotes the lambda closure: for a given set of functionals $\mathcal{S}$, $\lambda(\mathcal{S})$ is the set of functionals denoted by simply-typed lambda-terms using constants in $\mathcal{S}$; $\mathcal{S}_2$ is the restriction of $\mathcal{S}$ to second-order functionals. Type inference for $\sem{\st}$ is fully automatic and can be performed in time cubic in the size of the analyzed program. However the above characterization has two main weaknesses:
\begin{itemize}
\item It is not complete: As $\sem{\st} \subsetneq \BFF$, the typed language alone is not complete for $\BFF$ and a lambda closure (\ie, $\lambda(\mathcal{S})$) of functionals computed by typable and terminating programs is required to ensure completeness. 
\item It is not tractable: the set $\sem{\st}$ relies on a termination assumption and it is unclear whether the characterization still holds for a decidable or, for that matter, tractable termination technique.
\end{itemize}
\noindent 
Thus, providing a tractable, implicit, sound, and complete programming language for characterizing second-order polynomial time is still an open problem.

\subsection{Contributions}
Our paper provides the first solution to this problem, open for more than 20 years (\cite[Chapter~9]{IRK01}). To this end, we introduce a higher-order programming language and design a suitable typing discipline that address the two weaknesses described above. The lambda closure requirement for completeness is removed by designing a suitable programming language that consists of a layer of simply-typed terms that can perform calls to a layer of imperative and non-recursive procedures following a tier-based type discipline. This language allows for some restricted forms of procedure composition that are handled by the simply-typed terms and also allows for some restricted forms of oracle composition that are managed through the use of \emph{closures}, syntactic elements playing the role of first-order abstractions with free variables. The termination criterion is specified as a completeness-preserving instance, called $\SCPS$, of a variant of Size Change Termination~\cite{LJBA01} introduced in~\cite{BAL07} that can be checked in time quadratic in the size of the analyzed program. The main contributions of this paper are:
\begin{itemize}
\item A programming language in which typable ($\safe$) and terminating ($\SN$) programs capture exactly $\BFF$ (Theorem~\ref{thm:BFF}).
\item A restriction to lambda-free programs, called rank-$0$ programs, such that typable ($\safe_0$) and terminating ($\SN$) programs still capture exactly $\BFF$ (Theorem~\ref{thm:BFF0}); hence showing that lambda-abstraction only provides a syntactic relaxation,  and corresponds to a conservative extension in terms of computable functions.
\item A proof that type inference for $\safe$ is $\Ptime$-complete, and a type inference procedure running in time cubic in the program size for $\safe_0$ (Theorem~\ref{thm:tiprog}).
\item A simple termination criterion, called $\SCPS$, preserving soundness and completeness of the characterizations both for $\safe$ and for $\safe_0$ (Theorem~\ref{thm:sct}) that can be checked in  quadratic time.
\item A complete characterization of $\BFF$ in terms of typable ($\safe$) and terminating ($\SCPS$) programs (Theorem~\ref{thm:scti}) that captures strictly more programs (Figure~\ref{fig:ce}) than~\cite{HKMP20}, and is decidable in $\Ptime$-time.
\end{itemize}
\noindent 
The programming language under study comes with some usual restrictions on its expressive power, \ie, the typable programs it captures. Indeed, the studied characterizations of $\BFF$ are tractable/decidable whereas knowing whether a program computes a function in $\BFF$ is at least $\Sigma_0^2$-hard~\cite{H79}, hence not decidable. As a consequence, \emph{false negatives} (\ie, programs that are rejected by the characterizations but compute a function in $\BFF$) are unavoidable.

\subsection{Related work}
Several tools providing machine-independent characterizations of distinct complexity classes have been developed in the field of Implicit Computational Complexity (ICC). Most of these tools are restricted to the study of first-order complexity classes. Whereas  light logic approaches can deal with programs at higher types, their applications are restricted to first-order complexity classes such as $\FP$~\cite{Girard98,BT04,BM10}. Interpretation methods were extended to higher-order polynomials in~\cite{BL16} to study $\FP$ and adapted in~\cite{FHHP15} and~\cite{HP20} to characterize $\BFF$. However, these characterizations are not decidable as they require checking of second-order polynomial inequalities. 
\cite{CooKap89} and~\cite{IRK01} study characterizations of $\BFF$ in terms of a simple imperative programming language that enforces an explicit external bound on the size of oracle outputs within loops. The corresponding restriction is not implicit by nature and is impractical from a programming perspective as the size of oracle outputs cannot be predicted. In this paper, the bound is programmer friendly because it is implicit and it only constrains the size of the oracle input. 

\noindent 
The contributions of the paper are extensions of existing works:
\begin{itemize}
\item The soundness of the characterization relies strongly on the notion of \emph{continuation}, that fixes a given oracle (closure) for once in the imperative layer. If the oracle were allowed to be updated inside a while loop, depending on some local value, then the language would yield a class beyond $\BFF$, by computing exponential functions. Consequently, it would not be correct to think of $\safe_0$ as a simple extension of the language IMP of~\cite{W93} to higher-order because oracles need to be fixed once and for all in a call to a given imperative procedure. 
\item The characterization of $\BFF$ still holds in the absence of lambda-abstraction as a basic construct of the proposed programming language, in particular that completeness does not rely on lambda-abstractions apart from closures.
For that purpose, we consider an alternative characterization of $\BFF$ that was introduced in~\cite{KS19} using the second-order restriction $PV^{\omega}_2$ of the class $PV^{\omega}$ from~\cite{CU93}. Note that this characterization is not implicit as it contains explicit bounds.
This is an important improvement over~\cite{HKMP20} and~\cite{KS18}, both of which required external lambda-closure.
\item The type system is designed so that each procedure is typed exactly once. Types are not unique, but this does not prevent type inference from being polytime, as exhibiting one type is sufficient. The tractability of type inference is obtained by combining the tractability of type inference in the tier-based layer and in the simply-typed layer~\cite{M04}.
\item The particular choice of the termination criterion $\SCPS$ was made to show that termination can be specified as a tractable/feasible criterion  while preserving completeness. This is also a new result. $\SCPS$ may include nested loops (as described in~\cite{BAL07}) and can be replaced by any termination criterion capturing the programs used in the proof of completeness (Examples~\ref{ex:add},~\ref{ex:tm}, and~\ref{ex:it}). $\SCPS$ was chosen for its tractability, but not only: the $\SCP$ criterion of~\cite{BAL07} ensures termination by using an error state which breaks the control flow.
But this control-flow escape would damage the non-interference property needed for tier-based typing to guarantee time complexity bounds.
\end{itemize}

\noindent 
This paper is an extended and revised version of the paper~\cite{HKMP22} presented at Foundations of Software Science and Computation Structures - 25th                International Conference (FoSSaCS 2022), including complete proofs.

\subsection{Leading example} The program ${\tt ce}$, as in counterexample, of Figure~\ref{fig:ce}  will be our leading example, as it computes a function known to be in $\BFF -\sem{\st}$. This function was used by Kapron and Steinberg in~\cite{KS18} as a counterexample for showing that $\sem{\st}$ is not equal to $\BFF$. This program will be shown to be in $\safe_0$ and, consequently, in $\safe$ and to terminate with $\SCPS$.

\begin{figure} 
\lstset{language=,
morekeywords={declare, in, while, return, call, if, else, skip, @, var, box},
mathescape, frame=single, framesep=0pt, basicstyle=\ttfamily\small, lineskip=1pt, rulesepcolor=\color{gray!30},backgroundcolor=\color{gray!10},linewidth=\textwidth}
\newcommand*{\AddNotecopy}[4]{%
    \begin{tikzpicture}[overlay, remember picture]
        \draw [decoration={brace,amplitude=0.5em},decorate,ultra thick,black!60]
            ($(#3)!([yshift=1.5ex]#1)!($(#3)-(0,1)$)$) --  
            ($(#3)!(#2)!($(#3)-(0,1)$)$)
                node [align=left, text width=3.5cm, pos=0.5, anchor=west] {\quad #4};
    \end{tikzpicture}
}

\begin{lstlisting}
box $[$X,y$]$ in $\hspace*{112mm}$ $\tikzmark{listing-4-end}$
  declare KS(X$_1$, X$_2$, v) {$\hspace*{4.9cm}$ $\tikzmark{yerk}$ $\hspace*{0.5cm}$ $\tikzmark{listing-1-end}$
          var u,z$\sep$
        $\hspace*{4.5cm}$ $\tikzmark{truc1}$  $\tikzmark{momoche}$
          u $\asg$ X$_1$($\vide \upharpoonright \vide$)$\sep$
        $\tikzmark{truc2}$  
          z $\asg$ $\vide$$\sep$ $\tikzmark{truc3}$
          while (v != $\vide$) {
              v $\asg$ pred(v)$\sep$
              z $\asg$ X$_2$(z $\upharpoonright$ u)
          }$\tikzmark{truc4}$
          return z
  } $\tikzmark{listing-3-end}$
       $\tikzmark{listing-2-end}$   
  in call KS({x $\to$ X $\MVAt$ x}, {x $\to$ X $\MVAt$ (X $\MVAt$ x)}, y) $\tikzmark{listing-7-end}$
    $\tikzmark{listing-5-end}$     
 \end{lstlisting}
\AddNotecopy{listing-4-end}{listing-5-end}{listing-4-end}{{\tt ce}}
\AddNote{listing-1-end}{listing-3-end}{listing-7-end}{Procedure {\tt p}$_{\tt KS}$}
\AddNote{listing-2-end}{listing-5-end}{listing-7-end}{Term \texttt{t}}
\AddNotecopy{truc3}{truc4}{momoche}{Statement \texttt{st'}}
\AddNotecopy{truc1}{truc2}{momoche}{Statement \texttt{st}}
\caption{Program {\tt ce}}\label{fig:ce}
\end{figure}

Program ${\tt ce}$ computes a functional over the set of words  $\W$. 
When the boxed variables $\X$ and $\y$ in $\boite{\X,\y}$ are fed with the inputs $f \in \W \to \W$ and $w \in \W$, respectively, the term $\term$ calls the procedure ${\tt KS}$, named after Kapron-Steinberg. In ${\tt KS}$, variables $\X_1$ and $\X_2$ are substituted with closures $\{\x \to f\ \MVAt\ \x\}$ \ie, function $f$, and $\{\x \to f\ \MVAt\ (f\ \MVAt\ \x)\}$ \ie, function $f \circ f$, respectively. Variable $\vv$ is substituted with value $w$.
The binary operator $\upharpoonright$ truncates and pads the size of its first operand to the size of its second operand plus $1$.
 Procedure ${\tt KS}$ computes $\size{w}$ bounded iterations of $f \circ f$ through the execution of the assignment $\z \asg \X_2(\z \upharpoonright \uu)$. The bound on the output size of each iteration is computed by the first assignment $\uu \asg \X_1(\vide \upharpoonright \vide)$ of ${\tt KS}$ and is equal to $f(1)$, that is, $f(\sem{\vide \upharpoonright \vide})$, with $\sem{\vide \upharpoonright \vide}=1$; $\sem{\e}$ being the result of evaluating the expression $\e$. In any iteration of $f \circ f$, the internal call to $f$ is not bounded. Hence, there might be up to $\size{w}$ increases for the input of the outer call. Such a behavior is forbidden in $\sem{\st}$.

\section{A second-order language with imperative procedures}
\label{s:prog}
The syntax and semantics of the programming language designed to capture the complexity class $\BFF$ are introduced in this section. Programs of this language consist in second-order terms in which imperative procedures are declared and called. These procedures have no global variables, are not recursive, and their parameters can be of order 1 (oracles) or 0. Local variables inside a procedure can only be of order 0. 
Oracles are in read-only mode: they cannot be declared and, hence, modified inside a procedure. Oracles can only be composed at the term level through the use of closures, first-order abstractions that can be passed as parameters in a procedure call.

\subsection{Syntax}
Let $\overline{e}$ denote a possibly empty tuple of $n$ elements  $e_1,\ldots,e_n$, where $n$ is given by the context. Let $\size{\overline{e}}$ denote the length of tuple $\overline{e}$, \ie, $\size{\overline{e}}\triangleq n$. Let $\pi_i$, $0<i \leq \size{\overline{e}}$, denote the projectors on tuples, \ie, $\pi_i(\overline{e})\triangleq e_i$. Let $[e]$ denote some optional element $e$.

Let $\Var$ be a set of variables that can be split into three disjoint sets $\Var = \Var_0 \uplus\Var_1 \uplus \Var_{\geq 2}$. The type-$0$ variables in $\Var_0$ will be denoted by lower case letters $\x,\y,\ldots$ and the type-1 variables in $\Var_1$ will be denoted by upper case letters $\X,\Y,\ldots$ Variables in $\Var$ of arbitrary type will be denoted by letters $\va,\vb,\va_1,\va_2,\ldots$.

Let  $\Operator$ be a set of operators $\op$ of fixed arity $ar(\op)$ that will be used both in infix and prefix notations for notational convenience and that are always fully applied, \ie, applied to a number $ar(\op)$ of operands.

The programs are defined by the grammar of Figure~\ref{fig:synt} with keywords spelled out in bold typewriter font. A program is either a term $\term$,  a \emph{procedure declaration} $\letin{{\procedure}} \prog $, or the declaration of a \emph{boxed variable} $\va$, called \emph{box}, followed by a program: $\boite{\va}{\prog}$. Boxed variables will represent the program inputs.

When we refer to a \emph{type-$i$} syntactic element $e$ (a variable, an expression, a term, ...), for $i \in \mathbb{N}$, we implicitly assume that the element $e$ denotes some function of order $i$ over words as basic type.
For example, for a term $\term$, being type-$0$ means that $\term$ is simply-typed and its type does not contain an arrow. This notion will be formally defined in Section~\ref{s:type}.

In Figure~\ref{fig:synt}, there are three constructor/destructor pairs for abstraction and application; each of them playing a distinct r\^ole:
\begin{itemize}
\item $\lambda \va.\term$ and $\term \MVAt {\term}$ are the standard abstraction and application on terms.
\item The application of a type-1 variable $\X$ within a statement is called an \emph{oracle call}, written $\X(\e_1 \!\upharpoonright \e_2)$, where $\e_1$ is called the \emph{input data}, $\e_2$ is called the \emph{input bound}, and $\e_1 \!\upharpoonright \e_2$ is called the \emph{input}. The corresponding abstraction is called a \emph{closure}, a map of the shape $\{\x \to \term\}$, where the term $\term$ may contain free variables.
Closures will be restricted to be of type-$1$ by requiring $\term$ to be of type-$0$.
\item  A procedure declaration $\procname(\overline{\X},\overline{\x})\{[\var \overline{\y}\sep\!]\ \cmd  \ \ret \x\}$ is an abstraction that computes type-2 functions taking type-$1$ and type-$0$ inputs ($\overline{\X}$ and $\overline{\x}$, respectively) as parameters and returning a type-$0$ output $\x$. The \emph{procedure calls} of the shape $\callcc\ \procname(\overline{\clos},\overline{\term})$ are the corresponding applications and take closures as type-$1$ inputs and type-$0$ terms as type-$0$ inputs.
\end{itemize}

\begin{figure*}
\centering
$$ \begin{array}{llll}
\text{Type-0 var.} \qquad \qquad & \x
  & \in &\Var_0  \\
\text{Type-1 var.} \qquad \qquad & \X
  & \in &\Var_1  \\
\text{Variables} \qquad \qquad & \va
  & \in &\Var = \Var_0 \uplus \Var_1 \uplus \Var_{\geq 2}\\
\text{Procedure Names} \qquad & \procname
  & \in &\mathbb{P}\\
\text{Operators} \qquad \qquad & \op\ 
  & \in &\Operator  \\
\text{Expressions} \qquad \qquad & \e
  & \rgl &\x
   \ |\ \op(\overline{\e})  \ | \ \X (\e \!\upharpoonright \e) \\
\text{Statements} &  \cmd
    & \rgl &\skp   \ | \  \x  \asg \e   \ | \  \cmd \sep \cmd  \ |\ \ifa (\e) \{\cmd\} \elsea \{\cmd\} \\
   & & &
\ |\ \while(\e)\{ \cmd \}  \\
\text{Procedures} &\procedure
    & \rgl & \procname(\overline{\X},\overline{\x})\{[\var \overline{\x}\sep\!]\ \cmd  \ \ret \x\} \\
\text{Terms} \qquad \qquad & \term
  & \rgl & \va \ | \ \lambda \va.\term \ | \  \term \MVAt {\term} \ | \ \callcc\ \procname(\overline{\clos},\overline{\term}) \\
\text{Closures} \qquad \qquad & \clos
  & \rgl &\{  \x \to \term\} \\
\text{Programs} \qquad \qquad & \prog\ 
  & \rgl &  \term \ | \ \letin{\overline{\procedure}} \prog \ | \ \boite{\overline{\va}}{\prog}
 \end{array}$$
\caption{Syntax of type-2 programs}
\label{fig:synt}

\end{figure*}

\noindent 
For some syntactic element $e$ of the language, let $\FV(e) \subseteq \Var$ be the set of all variables occurring in $e$. A variable is free if it is not under the scope of an abstraction and it is not boxed. A program is \emph{closed} if it has no free variable. 

For a given procedure declaration $\procedure=\procname(\overline{\X},\overline{\x})\{[\var \overline{\y}\sep\!]\ \cmd  \ \ret \x \}$, define the \emph{procedure name} of $\procedure$ as $\n(\procedure)\triangleq \procname \in \mathbb{P}$, with $\mathbb{P}$ being a set of procedure names. Define also $\stt(\procedure)\triangleq \cmd $, $\local(\procedure)\triangleq\{\overline{\y}\}$, and $\param(\procedure) \triangleq \{\overline{\X},\overline{\x}\}$. $\stt(\procedure)$ is called the \emph{body} of procedure $\procedure$. The variables in $\local(\procedure)$ are called \emph{local variables} and the variables in $\param(\procedure)$ are called \emph{parameters}.

We will assume that for all programs the following well-formedness conditions hold:
there are no name clashes, that is two procedures should not share the same name and, in a procedure, a local variable cannot have the same name as a parameter; there are no free variables in a given procedure, in other words, the only variables in procedures are parameters and local variables; any procedure call has a corresponding procedure declaration.

Throughout the paper, we will consider closed programs in \emph{normal form}. These consist of programs with no free variable that can be written as follows $$\boite{\overline{\X},\overline{\x}}{\letin{\overline{\procedure}} \term},$$ for some term $\term$.
In other words, all type-$1$ variables then all type-$0$ variables are boxed before procedures are declared.

\subsection{Operational semantics}
Let $\W = \Sigma^*$ be the set of words over a finite alphabet  $\Sigma$ such that $\{0,1\} \subseteq \Sigma$. The symbol $\emptyword$ denotes the empty word. The length of a word $\w$  
is denoted  $\size{\w}$. 
Given two words $\w$ and $\vi$ in $\W$, let $\vi.\w$ denote the concatenation of $\vi$ and $\w$. For a given symbol $a \in \Sigma$, let $a^n$ be defined inductively by $a^0 = \emptyword$ and $a^{n+1}=a.a^n$.
Let $\dord$ be the sub-word relation over $\W$, which is defined by $\vi \dord \w$, if $\exists \ui,  \ui' \in \W, \ \w = \ui.\vi.\ui'$.

For a given word $\w \in \W$ and an integer $n$, let $\w_{\upharpoonright n}$ be the word obtained by truncating $\w$ to its first $\min(n, \size{w})$ symbols and then padding with a word of the form $10^k$ to obtain a word of size exactly $n+1$. For example, $1001_{\!\upharpoonright 0} = 1$, $1001_{\!\upharpoonright 1} = 11$, $1001_{\!\upharpoonright 2} = 101$, and $1001_{\!\upharpoonright 6} = 1001100$.
Define $\forall \vi,\w \in \W,\ \sem{\!\upharpoonright}(v,w) =  v_{\upharpoonright \size{w}}$.
Padding ensures that $\size{\sem{\!\upharpoonright}(v,w)}=\size{\w}+1$. The syntax of programs enforces that oracle calls are always performed on input data padded by the input bound and, consequently, oracle calls are always performed on input data whose size does not exceed the size of the input bound plus one.

We denote the set of total functions from set $A$ to set $B$ as $A\to B$ and the set of partial functions as $A\hookrightarrow B$.
For each operator $\op$ of arity $n$, a total function $\sem{\op} \in \W^{n} \to \W$ is defined. Constants may be viewed as operators of arity zero. We define two classes of operators called neutral and positive depending on the total function they compute. This categorization of operators will be used by our type system as the admissible types for operators will depend on their category.

\begin{defi}[Neutral and positive operators]\label{def:np} For an $n$-ary operator $\op$ that computes the total function $\sem{\op} \in \W^n \to \W$,
\begin{itemize}
\item $\op$ is \emph{neutral} if
$\sem{\op}$ is constant (\ie, $ar(\op)=0$),
$\sem{\op}\in\W^{n} \to \{0,1\}$ is a predicate,
 or $\forall \overline{\w} \in \W^{n},\ \exists i  \leq n$, $
\sem{\op}(\overline{\w}) \dord {\w_i}
$;
\item $\op$ is \emph{positive} if $\exists c_{\op} \in \mathbb{N}$ s.t.: $
\forall \overline{\w} \in \W^{n},\ \size{\sem{\op}(\overline{\w})}  \leq \max_{1 \leq i \leq n} \size{\w_i} + c_{\op}
$.
\end{itemize}
\end{defi}
\noindent 
As neutral operators are always positive, in the sequel, we reserve the name positive for those operators that are positive but not neutral.

In what follows, let $f,g,\ldots$ denote total functions from words to words.
A \emph{store} $\store$ consists of the disjoint union of a finite partial map $\store_0$ from $\Var_0$ to $\W$ and a finite partial map $\store_1$ from $\Var_1$ to total functions in $\W \to \W$.
Let $dom(\store)$ be the domain of the store $\store$. Let $\store[\x \leftarrow \w]$ denote the store $\store'$ satisfying $\store'(b)= \store(b)$, for all $b \neq \x$, and $\store'(\x)=\w$. This notation is extended naturally to type-1 variables $\store[\X \leftarrow f]$ and to sequences of distinct variables  $\store[\overline{\x} \leftarrow \overline{\w},\overline{\X} \leftarrow \overline{f}]$. Finally, let $\store_\emptyset$ denote the empty store.

Let $\downarrow$ denote the standard big-step call-by-name reduction relation on terms defined by: if $\term_1 \downarrow \lambda \va.\term$ and $\term \{\term_2/\va\} \downarrow v$ then $\term_1 \MVAt \term_2 \downarrow v$, where $\{\term_2/\va\}$ is the standard substitution and where $v$ can be a type-0 variable $\x$, a lambda-abstraction $\lambda \va.\term$, a type-1 variable application $\X \MVAt \term$, or a procedure call $\callcc\ \procname(\overline{\clos},\overline{\term})$.

A \emph{continuation} is a map $\phi$ from $\Var_1$ to the set of {Closures}, \ie,  $\phi(\X)=\{\x \to \term\}$ for some type-1 variable $\X$, some type-0 variable $\x$, and some type-0 term $\term$. Let $\overline{\X} \mapsto \overline{\clos}$ with $\size{\overline{\X}}=\size{\overline{\clos}}$, be a notation for the continuation mapping each $\X_i \in \Var_1$ to the closure $\clos_i$. 

Given a set of procedures $\sigma$, a store $\store$, and a continuation $\phi$, we define three distinct kinds of judgments: $(\sigma,\store,\phi,\e) \toexp  \w$ for expressions, $(\sigma,\store,\phi,\cmd) \tost \store'$ for statements, and $(\sigma,\store,\prog)\toenv \w$ for programs.
\begin{itemize}
\item The judgment $(\sigma,\store,\phi,\e) \toexp  \w$ means that the expression $\e$ evaluates to the word $\w \in \W$ with respect to the set of procedure declarations $\sigma$, the store $\store$ and the continuation $\phi$.
\item The judgment $(\sigma,\store,\phi,\cmd) \tost \store'$ expresses that, under the set of procedure declarations $\sigma$, the store $\store$ and the continuation $\phi$, the statement $\cmd$ terminates with resulting store $\store'$.
\item The judgment $(\sigma,\store,\prog)\toenv \w$ means that the tuple $(\sigma,\store,\prog)$ consisting of a set $\sigma$ of procedure declarations, a store $\store$, and a program $\prog$ maps deterministically to a word $\w \in \W$.
\end{itemize}
Given three sequences of words $\overline{\w}=\w_1,\ldots,\w_n$, expressions $\overline{\e}=\e_1,\ldots,\e_n$, and terms $\overline{\term}=\term_1,\ldots,\term_n$ of the same length $n$, we write $(\sigma,\store,\phi,\overline{\e}) \toexp  \overline{\w}$, if $\forall i \leq n$, $(\sigma,\store,\phi,\e_i) \toexp  \w_i$ and we write $(\sigma,\store,\overline{\term})\toenv \overline{\w}$, if $\forall i \leq n$, $(\sigma,\store,\term_i)\toenv \w_i$.

\begin{figure*} 
\begin{subfigure}[t]{\textwidth}
\centering
\begin{prooftree}
\hypo{\phantom{ \Imp \x  \asg \e\to } }
\infer1[(Var)]{(\sigma,\store,\phi \Imp \x) \toexp \store(\x)}
\end{prooftree}
\quad 
\begin{prooftree}
\hypo{(\sigma,\store,\phi,\overline{\e}) \toexp \overline{\w}}
\infer1[(Op)]{(\sigma,\store,\phi \Imp \op(\overline{\e})) \toexp \sem{\op}(\overline{\w})}
\end{prooftree}
\\[10pt]
\scalebox{0.93}{
\begin{prooftree}
\hypo{(\sigma,\store,\phi \Imp \e_1) \toexp \vi \quad (\sigma,\store,\phi \Imp \e_2) \toexp \ui \quad \phi(\X)=\{\x \to \term\} \quad (\sigma,\store[\x \leftarrow  \sem{\!\upharpoonright}(\vi,\ui)],\term ) \toenv \w }
\infer1[(Or)]{(\sigma,\store,\phi \Imp \X(\e_1\!\upharpoonright \e_2)) \toexp \w}
\end{prooftree}
}
\vspace{\baselineskip}
\caption{Expressions}
\vspace{\baselineskip}
\label{sfig:exp}
\end{subfigure}
\begin{subfigure}[t]{\textwidth}
\centering
\begin{prooftree}
\hypo{\phantom{(\sigma,\store,\phi \Imp \cmd_1) \tost \store' }}
\infer1[(Skip)]{(\sigma,\store,\phi \Imp \skp) \tost \store}
\end{prooftree}
\quad
\begin{prooftree}
\hypo{(\sigma,\store,\phi \Imp \cmd_1) \tost \store' }
\hypo{(\sigma,\store',\phi \Imp \cmd_2) \tost \store''}
\infer2[(Seq)]{(\sigma,\store,\phi \Imp \cmd_1 \sep \cmd_2) \tost \store''}
\end{prooftree}
\\[10pt]
\begin{prooftree}
\hypo{(\sigma,\store,\phi \Imp \e) \toexp \w }
\infer1[(Asg)]{(\sigma,\store,\phi \Imp \x  \asg \e) \tost \store[\x  \leftarrow \w]}
\end{prooftree}
\\[10pt]
\begin{prooftree}
\hypo{(\sigma,\store,\phi \Imp \e) \toexp \w}
\hypo{(\sigma,\store,\phi \Imp \cmd_\w) \tost \store'}
\hypo{\w\in \{\false,\true\} } 
\infer3[(Cond)]{(\sigma,\store,\phi \Imp \ifa (\e) \{\cmd_{\true}\} \elsea \{\cmd_{\false}\} ) \tost \store'}
\end{prooftree}
\\[10pt]
\begin{prooftree}
\hypo{(\sigma,\store,\phi \Imp \e) \toexp\false} 
\infer1[(Wh$_0$)]{(\sigma,\store,\phi \Imp \while (\e) \{\cmd\}) \tost \store}
\end{prooftree}
\\[10pt] 
\begin{prooftree}
\hypo{(\sigma,\store,\phi \Imp \e) \toexp \true}
\hypo{(\sigma,\store,\phi \Imp \cmd \sep  \while (\e) \{\cmd\}) \tost \store'}
\infer2[(Wh$_1$)]{(\sigma,\store,\phi \Imp  \while (\e) \{\cmd\}) \tost \store'}
\end{prooftree}
\vspace{\baselineskip}
\caption{Statements}
\vspace{\baselineskip}
\label{sfig:st}
\end{subfigure}

\begin{subfigure}[t]{\textwidth}
\centering
\begin{prooftree}
\hypo{\term \downarrow \x}
\hypo{\x \in \dom(\store)}
\infer2[(TVar)]{(\sigma,\store,\term) \toenv \store(\x)}
\end{prooftree}
\quad 
\begin{prooftree}
\hypo{\term \downarrow \X \MVAt \term_1}
\hypo{(\sigma,\store,\term_1 )\toenv \w}
\hypo{\X\in\dom(\store)}
\infer3[(OA)]{(\sigma,\store,\term) \toenv \store(\X)(\w)}
\end{prooftree}
\\[10pt]
\begin{prooftree}
\hypo{\term \downarrow \callcc\ \procname(\overline{\clos},\overline{\term_1})}
\hypo{(\sigma,\store,\overline{\term_1}) \toenv \overline{\w}}
\hypo{(\sigma,\store[\overline{\x} \leftarrow \overline{\w},\overline{\y} \leftarrow \overline{\emptyword}],\overline{\X} \mapsto \overline{\clos},\cmd ) \tost \store' }
\infer3[(Call)]{(\sigma \cup \{\procname(\overline{\X},\overline{\x})\{\var \overline{\y}\sep\!\ \cmd  \ \ret \z\} \},\store  \Imp {\term}) \toenv \store'(\z)}
\end{prooftree}
\vspace{\baselineskip}
\caption{Type-0 terms}
\vspace{\baselineskip}
\label{sfig:term}
\end{subfigure}
\begin{subfigure}[t]{\textwidth}
\centering
\begin{prooftree}
\hypo{(\sigma\cup\{\procedure\},\store,\prog) \toenv \w}
\infer1[(Dec)]{(\sigma,\store,\letin{{\procedure}} \prog) \toenv \w}
\end{prooftree}    
\quad \quad
\begin{prooftree}
\hypo{(\sigma,\store,\prog) \toenv \w}
\hypo{\va \in \dom(\store)}
\infer2[(Box)]{(\sigma,\store,\boite{\va}{\prog}) \toenv \w}
\end{prooftree}
\vspace{\baselineskip}
\caption{Programs}
\vspace{\baselineskip}
\label{sfig:prog}
\end{subfigure}
\caption{Big step operational semantics}
\label{fig:Com}
\end{figure*}
The big-step operational semantics of the language is described in Figure~\ref{fig:Com}.
One important point to stress is that closures cannot be nested and are set once and for all in a procedure call: they cannot be modified inside a procedure body.
It should also be noted that procedures do not have side effects as the modifications to the store are only overlaid and not carried over in the rest of the program.
In a closed program that respects well-formedness conditions, commuting $\boite{.}$ and $\letin{.}$ preserves semantics.
Hence for any closed program, there exists a semantically equivalent program in normal form.

A closed program in normal form $\prog= \boite{\overline{\X} ,\overline{\x}}{\letin{\overline{\procedure}} \term}$ computes the second-order partial functional $\sem{\prog} \in (\W \to \W)^{k} \to \W^{l} \hookrightarrow \W,$ defined by:
$$\sem{\prog}(f_1, \ldots, f_k,\w_1, \ldots, \w_l)=w \text{ iff } (\emptyset, \store_\emptyset[{\x}_1 \leftarrow {\w}_1, \ldots,{\X}_1 \leftarrow {f}_1 \ldots, \X_l \leftarrow f_l],\prog) \toenv  \w.$$

In the special case where $\sem{\prog}$ is a total function, the program $\prog$ is said to be terminating (strongly normalizing). We will denote by $\SN$ the set of terminating programs. For a given set of programs $S$, let $\sem{S}$ denote the set of functions computed by programs in $S$. Formally, $\sem{S}=\{ \sem{\prog} \ | \ \prog \in S\}$. For example, $\sem{\SN}$ is the set of total second-order functions computed by terminating programs.

\begin{exa}
Consider the program {\tt ce} provided in Figure~\ref{fig:ce}, where:
$$
\sem{\vide}() = \emptyword \in \W, \quad
\sem{\meq{}}(\w, \vi) = 
   \begin{cases}
     1 &\text{if }\vi\neq\w  \\
         0 &\text{otherwise} 
   \end{cases},      \quad          
\sem{\mpred}(\vi)   =
     \begin{cases}
       \emptyword  &\text{if\ }\vi=\emptyword \\ 
      \ui  &\text{if\ } \vi=a.\ui,\ a \in \Sigma
      \end{cases}.    
$$

\noindent 
Program {\tt ce} is in normal form and computes the second-order functional $F \in (\W \to \W) \to \W \to \W$ defined by: $\forall f \in \W \to \W, \forall \w \in \W,\ F(f)(\w) = F_{|\w|}(f),$ where $F_n$ is defined recursively as 
$$\begin{array}{rcl}
  F_0(f)&=&\emptyword\\
  F_{n+1}(f) &=&(f \circ f)(\sem{\upharpoonright}(F_n(f),f(1))= (f \circ f)(F_n(f)_{\upharpoonright |f(1)|}).
\end{array}$$
That is a function that composes the input function $2|\w|$ times $f$ while restricting its input to a fixed size $\size{f(1)}$ every other iteration.  Indeed, $\sem{\vide}()=\emptyword$ and $\sem{\upharpoonright}(\emptyword,\emptyword)=\emptyword_{\upharpoonright \size{\emptyword}}=1$. Consequently, the oracle bound $\uu$ in the oracle call $\X_2(\z \upharpoonright \uu)$ is bound to value $f(1)$ in the store by the statement $\uu \asg \X_1(\vide \upharpoonright \vide)$.

Observe that the operators $\vide$, $\meq{}$ and $\mpred{}$ are all neutral.  An example of positive operator can be given by the successor operators defined by 
$\sem{\msuc{i}}(\vi) = i.\vi, \ \text{for }i \in \{0,1\}$. These operators are positive since $\size{\sem{\msuc{i}}(\vi)}=\size{i.\vi}=\size{\vi}+1$.

The operational semantics of {\tt ce} is illustrated by the judgment provided in Figure~\ref{fig:red}, that represents a portion of the evaluation of the whole program on inputs $f$ and $\w$. 
\end{exa}

\begin{figure}
\begin{tikzpicture}
\node[rotate=-90] at (0,0) {\begin{minipage}{0.987\textheight}
\hrulefill

Let $ \procedure_{\mathtt{KS}} ={\tt KS}(\X_1,\X_2,{\vv})\{\var \uu,\z \sep \cmd \sep \cmd' \ \ret \z\}$, with $\cmd = \uu \asg \X_1(\vide \upharpoonright \vide)$,\vfill
$\store_1 = \store_\emptyset[\X \leftarrow f, \y \leftarrow \w,\uu \leftarrow \emptyword,\z \leftarrow \emptyword]$, \vfill
$\store_2(\z)=F_{\size{\w}}(f)$, \vfill 
and $\phi(\X_1)=\{\x \to \X \MVAt \x\}$,  $\phi(\X_2)=\{\x \to \X \MVAt (\X \MVAt \x)\}$ in:

\begin{center}
$$
\scalebox{0.775}{
\begin{prooftree}
\hypo{\store_\emptyset[\X \leftarrow f, \y \leftarrow \w](\y)=\w}
\infer1[(TVar)]{(\{\procedure_{\mathtt{KS}}\},\store_\emptyset[\X \leftarrow f, \y \leftarrow \w],\y) \toenv {\w}}
\hypo{\phi(\X_1)=\{\x \to\X \MVAt \x\}}
\hypo{ }
\infer1[(TVar)]{(\{\procedure_{\mathtt{KS}}\},\store_1[\x \leftarrow 1],\x )\toenv 1}
\infer1[(OA)]{(\{\procedure_{\mathtt{KS}}\},\store_1[\x \leftarrow \sem{\!\upharpoonright}(\emptyword,\emptyword)],\X \MVAt \x ) \toenv f(1) }
\infer2[(Or)]{(\{\procedure_{\mathtt{KS}}\},\store_1,\phi \Imp \X_1(\vide \upharpoonright \vide)) \toexp f(1) }
\infer1[(Asg)]{(\{\procedure_{\mathtt{KS}}\},\store_1,\phi \Imp \uu \asg \X_1(\vide \upharpoonright \vide)) \tost \store_1[\uu \leftarrow f(1)] }
\hypo{}
\ellipsis{}{}
\infer1[(Seq)]{(\{\procedure_{\mathtt{KS}}\},\store_1[\uu \leftarrow f(1)],\phi,\cmd') \tost \store_2}
\infer2[(Seq)]{(\{\procedure_{\mathtt{KS}}\},\store_1,\phi,\cmd \sep \cmd' ) \tost \store_2
}
\infer2[(Call)]{(\{\procedure_{\mathtt{KS}}\},\store_\emptyset[\X \leftarrow f, \y \leftarrow \w],\callcc\ {\tt KS}(\{\x \to \X \MVAt \x\},\{\x \to \X \MVAt (\X \MVAt \x)\},\y)) \toenv F_{\size{\w}}(f)}
\infer1[(Dec)]{(\emptyset,\store_\emptyset[\X \leftarrow f, \y \leftarrow \w],\letin{{\procedure}_{\mathtt{KS}}} \callcc\ {\tt KS}(\{\x \to \X \MVAt \x\},\{\x \to \X \MVAt (\X \MVAt \x)\},\y))  \toenv F_{\size{\w}}(f)}
\infer1[(Box)]{\ldots \phantom{\store}}
\infer1[(Box)]{(\emptyset,\store_\emptyset[\X \leftarrow f, \y \leftarrow \w],{\tt ce}) \toenv F_{\size{\w}}(f)}
\end{prooftree}
}
$$
\hrulefill
\end{center}
\centering
\captionof{figure}{Evaluation of the program \texttt{ce} of Figure~\ref{fig:ce}}\label{fig:red}
\end{minipage}};
\end{tikzpicture}
\end{figure}

\section{Type system}\label{s:type}
In this section, we introduce the typing discipline for our programming language.

\subsection{Tiers and typing environments}

Let $\TW$ be the type of words in $\W$. \emph{Simple types} over $\TW$ are defined inductively by
$\STW,\STW',\ldots \rgl \TW \ | \ \STW \to \STW.$
Let $\SST$ be the set of simple types over $\TW$. The order of a simple type in $\SST$ is defined inductively by:
$\order(\STW)=0$, if $\STW = \TW$, and $\order(\STW)=\max(1+\order(\STW_1),\order(\STW_2))$, if $\STW = \STW_1 \to \STW_2$.

\emph{Tiers} are elements of the totally ordered set $(\SL ,\ord)$, where $\SL =\{\tiera, \tierb, \tierc, \ldots\}$ is the set of natural numbers with order $\ord$. The strict order on $\SL$ will be denoted by $\ordst$ and the binary $\meet$ and $\join$ denote the usual operations. We use bold symbols $\sla,\slb,\ldots,\sla_1,\sla_2,\ldots$ to denote tier variables. For $n$ tiers $\{\sla_1,\ldots,\sla_n\}$, we also define the $n$-ary max and min: $\join_{i=1}^n \sla_i$ and $\meet_{i=1}^n \sla_i$.
 A \emph{first-order tier} is of the shape $\sla_1 \to \ldots \to \sla_{n} \to \sla'$, with $\sla_i,\sla' \in \SL$.

A \emph{simple typing environment} $\typenvs$ is a finite partial map from $\Var$ to $\SST$, which assigns simple types to variables.

A \emph{variable typing environment} $\typenv$ is a finite partial map from $\Var_0$ to $\SL $, which assigns single tiers to type-0 variables.

An \emph{operator typing environment} $\typop$ is a mapping that associates to some operator $\op$ and some tier $\sla \in \SL$ a set of first-order tiers $\typop(\op)(\sla)$ of the shape $\sla_1 \to \ldots \to \sla_{n} \to \sla'$ where $n$ is the arity of $\op$.

A \emph{procedure typing environment} $\typproc$ is a mapping that associates to each procedure $\procedure$ a pair $\langle\typenv,\overline{\sla}\rangle$ consisting of a variable typing environment $\typenv$ and a triplet of tiers $\overline{\sla}$. Let $\typproc_i \triangleq \pi_i(\typproc)$, $i \in \{1,2\}$.

Let $\dom(\typenv)$, $\dom(\typenvs)$, $\dom(\typop)$, and $\dom(\typproc)$ denote the sets of variables typed by $\typenv$ and $\typenvs$, the set of operators typed by $\typop$, and the set of procedures typed by $\typproc$, respectively.

For a procedure typing environment $\typproc$, it will be assumed that for every $\procedure \in \dom(\typproc)$, $\param(\procedure) \cup \local(\procedure)  \subseteq \dom(\typproc_1(\procedure))$.

While operator and procedure typing environments are global, \ie, defined for the whole program, variable typing environments are local, \ie, relative to the procedure under analysis. In a program typing judgment, the simple typing environment can be viewed as the typing environment for the main program.

\subsection{Typing judgments and type system} The typing discipline includes two distinct kinds of typing judgments:
\begin{itemize}
\item \emph{Procedure typing judgments} $\pbl e : (\sla,\sla_{in},\sla_{out})$, with  with  $\sla,\sla_{in},\sla_{out} \in \SL$, and $e$ an expression or a statement; 
\item \emph{Term typing judgments} $\pbla \prog : \STW$, with $\STW \in \SST$. 
\end{itemize}

\noindent 
 The  meaning of the procedure typing judgment is that the \emph{expression tier} (or \emph{statement tier}) is $\sla$, the \emph{innermost  tier} is $\sla_{in}$, and  the \emph{outermost tier} is $\sla_{out}$. 
The innermost (resp. outermost) tier is the tier of the innermost (resp. outermost) while loop guard where the expression or statement is located.
The meaning of term typing judgments is that the program $\prog$ is of simple type $\STW$ under the operator typing environment $\typop$, the procedure typing environment $\typproc$ and the simple typing environment $\typenvs$.

A program $\prog$ (or term $\term$) is of \emph{type-$i$}, if $\pbla \prog : \STW$ ($\pbla \term : \STW$) can be derived for some typing environments and type $\STW$ s.t. $\order(\STW)=i$.

The type system for the considered programming language is provided in Figure~\ref{fig:TS}. The typing rules for expressions and statements in Figure~\ref{sfig:tsproc} are the same as in~\cite{HKMP20}; The typing of procedures, terms, closures and programs in Figure~\ref{sfig:tsprog} is new.
A \emph{well-typed program} is a program that can be given the type $(\overline{\TW \to \TW}) \to \overline{\TW} \to \TW$, \ie, the judgment $\pbla {\prog} : (\overline{\TW \to \TW}) \to \overline{\TW} \to \TW$ can be derived for the environments $\typenvs,\typproc, \typop$. Consequently, a well-typed program is a type-$i$ program, for some $i \leq 2$, computing a functional.

Due to the rule (E-OP) of Figure~\ref{sfig:tsproc}, which allows several admissible types for operators, typing derivations are, in general, not unique. However, under  the assumption of existence, program types are unique for  fixed typing environments. The two typing rules for while loops (S-WH) and (S-WINIT) are mutually exclusive because of the non-overlapping requirements for $\sla_{out}$ in Figure~\ref{sfig:tsproc}. (S-WH) is the standard rule and updates the innermost tier with the tier of the while loop guard under consideration. (S-WINIT) is an initialization rule that allows the programmer to instantiate by default the main command with outermost tier $\tiera$ as it has no outermost while. It could be sacrificed for simplicity but at the price of a worse expressive power.

For a given typing judgment $j$, a \emph{typing derivation} $\pi \rightslice j$  is a tree whose root is the (procedure or term) typing judgment $j$ and whose children are obtained by applications of the typing rules of Figure~\ref{fig:TS}.  The name $\pi$ will be used alone whenever mentioning the root of a typing derivation is not explicitly needed. A typing sub-derivation of a typing derivation $\pi$ is a subtree of $\pi$.

\begin{figure} 
\begin{subfigure}{\textwidth}
\centering
\begin{prooftree}
\hypo{\typenv(\x)=\sla}
\infer1[(E-VAR)]{\pbl \x: (\sla, \sla_{in},\sla_{out})}
\end{prooftree}
\\[4pt]
\begin{prooftree}
\hypo{\sla_1 \to \cdots \to \sla_{n} \to \sla \in \Delta(\op)(\sla_{in})}
\hypo{\forall i \leq n,\ \pbl \e_i: (\sla_i,\sla_{in},\sla_{out})}
\infer2[(E-OP)]{\pbl  \op(\e_1,\ldots,\e_{n}): (\sla,\sla_{in},\sla_{out})}
\end{prooftree}
\\[4pt]
\begin{prooftree}
\hypo{\pbl \e_1:(\sla,\sla_{in},\sla_{out})}
\hypo{\pbl \e_2: (\sla_{out},\sla_{in},\sla_{out})}
\hypo{\sla \ordst  \sla_{in} \text{ and } \sla \ord \sla_{out}  }
\infer3[(E-OR)]{\pbl \X(\e_1 \!\upharpoonright \e_2) : ( \sla,\sla_{in},\sla_{out})}
\end{prooftree}
\\[4pt]
\begin{prooftree}[center=false]
\infer0[(S-SK)]{\pbl \skp\ : (\tiera,\sla_{in},\sla_{out})}
\end{prooftree}
\qquad
\begin{prooftree}[center=false]
\hypo{\pbl  \cmd\ : (\sla,\sla_{in},\sla_{out})}
\infer1[(S-SUB)]{\pbl  \cmd\ : (\sla\mathbf{+1},\sla_{in},\sla_{out})}
\end{prooftree}
\\[4pt]
\begin{prooftree}
\hypo{\pbl  \cmd_1 :(\sla,\sla_{in},\sla_{out})}
\hypo{\pbl \cmd_2:(\sla,\sla_{in},\sla_{out})}
\infer2[(S-SEQ)]{\pbl \cmd_1 \sep \cmd_2 \ :(\sla,\sla_{in},\sla_{out})}
\end{prooftree}
\\[4pt]
\begin{prooftree}
\hypo{\pbl \x: (\sla_1,\sla_{in},\sla_{out})}
\hypo{\pbl \e: (\sla_2,\sla_{in},\sla_{out})}
\hypo{\sla_1  \ord \sla_2}
\infer3[(S-ASG)]{\pbl \x \asg \e \ : (\sla_1,\sla_{in},\sla_{out})}
\end{prooftree}
\\[4pt]
\scalebox{0.9}{\begin{prooftree}
\hypo{\pbl \e: (\sla,\sla_{in},\sla_{out})}
\hypo{\pbl \cmd_{\true}: (\sla,\sla_{in},\sla_{out})}
\hypo{\pbl \cmd_{\false}:(\sla,\sla_{in},\sla_{out})}
\infer3[(S-CND)]{\pbl \ifa (\e) \{\cmd_{\true}\} \elsea \{\cmd_{\false}\}\ :(\sla,\sla_{in},\sla_{out})}
\end{prooftree}}
\\[4pt]
\begin{prooftree}
\hypo{\pbl \e: (\sla,\sla_{in},\sla)}
\hypo{\pbl \cmd : (\sla,\sla,\sla)}
\hypo{\tierb \ord \sla }
\infer3[(S-WINIT)]{\pbl  \while (\e) \{\cmd\}\ :(\sla,\sla_{in},\tiera)}
\end{prooftree}
\\[4pt]
\begin{prooftree}
\hypo{\pbl \e: (\sla,\sla_{in},\sla_{out})}
\hypo{\pbl \cmd : (\sla,\sla,\sla_{out})}
\hypo{\tierb \ord \sla \ord \sla_{out}}
\infer3[(S-WH)]{\pbl  \while (\e) \{\cmd\}\ :(\sla,\sla_{in},\sla_{out})}
\end{prooftree}
\vspace{\baselineskip}
\caption{Tier-based typing rules for expressions and statements (from \cite{HKMP20})}
\label{sfig:tsproc}
\end{subfigure}
\begin{subfigure}{\textwidth}
\vspace{\baselineskip}
\centering
\begin{prooftree}
\hypo{\pbla \overline{\X} : \overline{\TW \to \TW}}
\hypo{\pbla \overline{\x},\overline{\y} : \overline{\TW}}
\hypo{\pbla \x : {\TW}}
\infer3[(PR-DEC)]{\pbla \procname(\overline{\X},\overline{\x})\{[\var \overline{\y}\sep\!]\ \cmd  \ \ret \x\} : (\overline{\TW \to \TW}) \to \overline{\TW} \to \TW}
\end{prooftree}
\\[4pt]
\scalebox{0.9}{\begin{prooftree}
\hypo{\pbla  \procname(\overline{\X},\overline{\x})\{\ldots\} : (\overline{\TW \to \TW}) \to \overline{\TW} \to \TW}
\hypo{\pbla  \overline{\clos} : \overline{\TW \to \TW}}
\hypo{\pbla \overline{\term} : \overline{\TW}}
\infer3[(P-CALL)]{\pbla \callcc\ \procname(\overline{\clos},\overline{\term}) : \TW}
\end{prooftree}}
\\[4pt]
\begin{prooftree}
\hypo{\typenvs(\va) = \STW}
\infer1[(P-VAR)]{\pbla \va : \STW}
\end{prooftree}
\quad
\begin{prooftree}
\hypo{\typenvs \uplus \{\va : \STW\},\typproc,\typop \vdash \term :\STW' }
\infer1[(P-ABS)]{\pbla \lambda \va.{\term} : \STW \to \STW'}
\end{prooftree}
\\[4pt]
\begin{prooftree}
\hypo{\pbla \term_1 : \STW \to \STW'}
\hypo{\pbla \term_2 : \STW}
\infer2[(P-APP)]{\pbla \term_1 \MVAt {\term_2} : \STW'}
\end{prooftree}
\\[4pt]
\scalebox{0.95}{\begin{prooftree}
\hypo{\pbla \prog : {\STW}}
\hypo{\typenv,\typop \vdash \stt(\procedure) : (\sla,\sla_{in},\sla_{out})}
\hypo{\typproc(\procedure) = \langle \typenv, (\sla,\sla_{in},\sla_{out}) \rangle}
\infer3[(P-DEC)]{\pbla \letin{{\procedure}} \prog : \STW}
\end{prooftree}}
\\[4pt]
\begin{prooftree}
\hypo{\typenvs\uplus \{\x : \TW\} ,\typproc,\typop \vdash \term : \TW}
\infer1[(P-CLOS)]{\pbla \{\x \to \term\} : \TW \to \TW}
\end{prooftree}
\\[4pt]
\begin{prooftree}
\hypo{\typenvs\uplus\{\va:\STW\},\typproc,\typop \vdash \prog : {\STW'}}
\infer1[(P-BOX)]{\pbla \boite{\va}{\prog} : \STW \to \STW'}
\end{prooftree}
\vspace{\baselineskip}
\caption{Simple typing rules for procedures, terms, closures and programs}
\label{sfig:tsprog}
\end{subfigure}
\caption{Tier-based type system}
\label{fig:TS}
\end{figure}

Notice that the type system of Figure~\ref{fig:TS} is a strict extension of the type system of~\cite{HKMP20}, which mostly consists of the typing rules in Figure~\ref{sfig:tsproc}, for procedures.

\subsection{Safe programs}\label{ss:safe}
In this section, we restrict the set of admissible operators to prevent programs admitting exponential growth from being typable. A program satisfying such a restriction will be called \emph{safe}.
We first define a safe operator typing environment by restraining how neutral and positive operators can be used, in the same way as in~\cite{HKMP20}.
 
\begin{defi}[Safe operator typing environment]\label{sote}
An operator typing environment $\typop$ is \emph{safe} if for each $\op \in \dom(\typop)$ with non null arity $n$, 
$\op$ is neutral or positive,
$\sem{\op}$ is a polynomial time computable function,
and for each $\sla \in \SL$,
for each $\sla_1 \to \ldots \sla_{n} \to \sla' \in \typop(\op)(\sla)$, the two conditions below hold:
\begin{enumerate}
\item $\sla' \ord \mini_{i=1}^{n} \sla_i \ord \maxi_{i=1}^{n} \sla_i \ord \sla$,\label{premier}
\item  if $\op$ is a positive operator then $\sla' \ordst \sla$.\label{second}
\end{enumerate}
\end{defi}

\begin{exa}\label{ex:safeop}
Consider the operators $\meq{}$, $\mpred$, and $\msuc{i}$ discussed in Figure~\ref{fig:ce} and an operator typing environment $\Delta$ that is safe and such that $\meq{}$, $\mpred$, $\msuc{i}$ $\in \dom(\Delta)$.
We can set $\Delta(\meq{})(\tierb)=\{\tierb \to \tierb \to \tierb \} \cup \{\sla \to \sla' \to \tiera \ | \ \sla,\sla' \ord \tierb\}$, as $\meq{}$ is neutral. However $\tierb \to \tiera \to \tierb \notin \Delta(\meq{})(\tierb)$ as  it breaks Condition~(\ref{premier}) above  (\ie, $ \tierb \ord \meet(\tierb,  \tiera)$ is false).

We can also set $\Delta(\mpred)(\tierc)=\{\tierc \to \sla \ | \ \sla \ord \tierc \}\cup \{ \tierb \to \sla \ | \ \sla \ord \tierb \} \cup \{\tiera \to \tiera \}.$ We also have $\Delta(\msuc{i})(\tierb) =\{ \tierb \to \tiera ,\tiera \to \tiera\}.$ $\tierb \to \tierb \notin \Delta(\msuc{i})(\tierb)$ as $\msuc{i}$ is a positive operator and, due to Condition~(\ref{second}) above, the operator output tier has to be strictly smaller than $\tierb$. 
\end{exa}

A program $\prog$ is a \emph{safe program} if there exist a simple typing environment $\typenvs$, a procedure typing environment $\typproc$, and a safe operator typing environment such that $\prog$ is well-typed for these environments, \ie,  $\pbla \prog : (\overline{\TW \to \TW}) \to \overline{\TW} \to \TW$ can be derived. Let $\safe$ be the set of safe programs.

\begin{exa}\label{ex:type}
We consider the program ${\tt ce}$ of Figure~\ref{fig:ce}.
We define the operator typing environment $\typop$ by
\begin{align*}
\typop(\meq{})(\tierc) &\triangleq \{\tierb \to \tierb \to \tierb\},\\
\typop(\mpred{})(\tierb) &\triangleq \{\tierb \to \tierb\},\\
\typop(\vide)(\tierc) &\triangleq \{\tiera , \tierb\}.
\end{align*}
As the three operators $\meq{}$, $\mpred{}$, and $\vide$ are neutral, the environment $\typop$ is safe.

We define the simple typing environment $\typenvs$ by
$\typenvs(\uu) \triangleq \TW, \typop(\vv)\triangleq \TW, \typop(\z) \triangleq \TW, \typenvs(\X_1) \triangleq \TW \to \TW$, and $ \typenvs(\X_2) \triangleq \TW \to \TW$. We define the variable typing environment $\typenv$ by $\typenv(\uu) \triangleq \tierb, \typop(\vv)\triangleq \tierb, \typop(\z) \triangleq \tiera$. Finally, define the procedure typing environment $\typproc$ by $\typproc(\procedure_{\tt KS}) \triangleq \langle \typenv, (\tierb,\tierc,\tierb) \rangle$. Using the rules of Figure~\ref{fig:TS}, the following typing judgment can be derived $ \pbla {\tt ce} : ({\TW \to \TW}) \to {\TW} \to \TW .$

The full typing derivation is provided in Figure~\ref{fig:typex} where we sometimes use the notation $\pbla \overline{e} : \overline{\STW}$ as a shorthand notation for $\forall i \leq \size{\overline{e}},\ \pbla {e_i} : {\STW}$ under the assumption that $\size{\overline{e}}=\size{\overline{\STW}}$. The operator typing environment is safe and, consequently, ${\tt ce}$ is a safe program, \ie, ${\tt ce} \in \safe$.
\end{exa}

\begin{figure}
\begin{tikzpicture}
\node[rotate=-90] at (0,0) {\begin{minipage}{0.987\textheight}
\hrulefill
                
Let $\diamond_{\va_1, \ldots, \va_n}^{\STW_1,\ldots, \STW_n} \triangleq  \typenvs \uplus \{\va_1 : \STW_1,\ldots, \va_n : \STW_n\}, \Omega, \Delta$, for $n \geq 0$, and let $\TW^\TW \triangleq \TW \to \TW$ in:                            

$$
\scalebox{0.64}{
\begin{prooftree}
\hypo{}
\infer1[(P-VAR)]{\pblap \X_1,\X_2,\vv,\z,\uu : \overline{\TW^\TW},\overline{\TW}}
\infer1[(PR-DEC)]{\pblap  {\tt KS}(\X_1,\X_2,\vv)\ldots  : ({\TW^\TW})^2 \to {\TW} \to \TW}
\hypo{\pi_1'}
\ellipsis{}{}
\infer1[(P-VAR)]{\pblapp \X \MVAt (\X \MVAt \x) : \TW}
\infer1[(P-CLOS)]{\pblap  \{\x \to \X \MVAt (\X \MVAt \x)\} : {\TW^\TW}}
\hypo{\pi_1''}
\ellipsis{}{}
\infer1[(P-VAR)]{\pblapp   \X \MVAt \x : \TW}
\infer1[(P-CLOS)]{\pblap \{\x \to \X \MVAt \x\} : {\TW^\TW}}
\hypo{}
\infer1[(P-VAR)]{\pblap \y : {\TW}}
\infer4[(P-CALL)]{\pblap \callcc \ {\tt KS}(\{\x \to \X \MVAt \x\},\{\x \to \X \MVAt (\X \MVAt \x)\},\y) : {\TW}}
\hypo{\pi_2}
\ellipsis{}{}
\infer1[(S-SEQ)]{\typenv,\Delta \vdash \stt(\procedure_{\tt KS}) : (\tierb,\tierc,\tierb)}
\infer2[(P-DEC)]{\pblap\letin{\procedure_{\tt KS}}{\callcc \ {\tt KS}(\{\x \to \X \MVAt \x\},\{\x \to \X \MVAt (\X \MVAt \x)\},\y)} : \TW}
\infer1[(P-BOX)]{\diamond_\X^{\TW^\TW} \vdash\boite{\y}{\letin{\procedure_{\tt KS}}{\callcc \ {\tt KS}(\{\x \to \X \MVAt \x\},\{\x \to \X \MVAt (\X \MVAt \x)\},\y)}} : (\TW^\TW) \to \TW^\TW}
\infer1[(P-BOX)]{\pbla \boite{\X,\y}{\letin{\procedure_{\tt KS}}{\callcc \ {\tt KS}(\{\x \to \X \MVAt \x\},\{\x \to \X \MVAt (\X \MVAt \x)\},\y)}} : (\TW^\TW) \to \TW^\TW}
\end{prooftree}
}
$$

where $\procedure_{\tt KS} =\texttt{KS}(\X_1, \X_2, \vv)\{\uu \asg \X_1(\vide \upharpoonright \vide)\sep \z \asg \vide\sep  \while (\vv\ \meq{}\ \vide) \{\vv \asg \mpred{}(\vv)\sep \z \asg \X_2(\z \upharpoonright \uu)\}\}$, $\typproc(\procedure_{\tt KS}) = \langle \typenv, (\tierb,\tierc,\tierb) \rangle$, and

$$
\pi_2=\scalebox{0.68}{
\begin{prooftree}
\hypo{\typenv(\uu)=\tierb}
\infer1[(E-VAR)]{\typenv,\typop \vdash \uu: (\tierb,\tierc,\tierb)}
\hypo{ \tierb \in \typop(\vide)(\tierc)}
\infer1[(E-OP)]{\typenv,\typop \vdash \vide : (\tierb,\tierc,\tierb)}
\infer1[(E-OR)]{\typenv,\typop \vdash \X_1(\vide \upharpoonright \vide): (\tierb,\tierc,\tierb)}
\infer2[(S-ASG)]{\typenv,\typop \vdash \uu \asg \X_1(\vide \upharpoonright \vide) : (\tierb,\tierc,\tierb)   }
\hypo{}
\infer[no rule]1{\typenv(\z)=\tiera}
\infer1[(E-VAR)]{\typenv,\typop \vdash \z : (\tiera,\tierc,\tierb)}
\hypo{}
\infer[no rule]1{ \tiera \in \typop(\vide)(\tierc)}
\infer1[(E-OP)]{\typenv,\typop \vdash \vide: (\tiera,\tierc,\tierb)}
\infer2[(S-ASG)]{\typenv,\typop \vdash \z \asg \vide : (\tiera,\tierc,\tierb)  }
\infer1[(S-SUB)]{\typenv,\typop \vdash \z \asg \vide : (\tierb,\tierc,\tierb)  }
\hypo{\pi_2'}
\ellipsis{}{}
\infer1{\typenv,\typop \vdash\while (\vv\ \meq{}\ \vide) \{\ldots\} : (\tierb,\tierc,\tierb) }
\infer2[(S-SEQ)]{\typenv,\typop \vdash \z \asg \vide\sep  \while (\vv\ \meq{}\ \vide) \{\vv \asg \mpred{}(\vv)\sep \z \asg \X_2(\z \upharpoonright \uu)\} : (\tierb,\tierc,\tierb)   }
\infer2[(S-WH)]{\pi_2 \rightslice \typenv,\typop \vdash \uu \asg \X_1(\vide \upharpoonright \vide)\sep \z \asg \vide\sep  \while (\vv\ \meq{}\ \vide) \{\vv \asg \mpred{}(\vv)\sep \z \asg \X_2(\z \upharpoonright \uu)\} : (\tierb,\tierc,\tierb)   }
\end{prooftree}
}
$$

$$
\pi_2'=
\scalebox{0.69}{
\begin{prooftree}
\hypo{\tierb \to \tierb \to \tierb \in \typop(\meq{})(\tierc)}
\hypo{\typenv(\vv)=\tierb}
\infer1[(E-VAR)]{\typenv,\typop \vdash  \vv  : (\tierb,\tierc,\tierb)}
\hypo{ \tierb \in \typop(\vide)(\tierc)}
\infer1[(E-OP)]{\typenv,\typop \vdash  \vide  : (\tierb,\tierc,\tierb)}
\infer3[(E-OP)]{\typenv,\typop \vdash  \vv\ \meq{}\ \vide : (\tierb,\tierc,\tierb) }
\hypo{\pi_2''}
\ellipsis{}{}
\infer1[(S-SEQ)]{\typenv,\typop \vdash \vv \asg \mpred{}(\vv)\sep \z \asg \X_2(\z \upharpoonright \uu) : (\tierb,\tierb,\tierb) }
\infer2[(S-WH)]{\pi_2' \rightslice \typenv,\typop \vdash\while (\vv\ \meq{}\ \vide) \{\vv \asg \mpred{}(\vv)\sep \z \asg \X_2(\z \upharpoonright \uu)\} : (\tierb,\tierc,\tierb)}
\end{prooftree}
}
$$

$$
\pi_2''=\scalebox{0.67}{
\begin{prooftree}
\hypo{\typenv(\vv)=\tierb}
\infer1[(E-VAR)]{\typenv,\typop \vdash \vv: (\tierb,\tierb,\tierb) }
\hypo{\tierb \to \tierb \in \typop(\mpred{})(\tierb)}
\hypo{\typenv(\vv)=\tierb}
\infer1[(E-VAR)]{\typenv,\typop \vdash \vv: (\tierb,\tierb,\tierb) }
\infer2[(E-OP)]{\typenv,\typop \vdash \mpred{}(\vv): (\tierb,\tierb,\tierb) }
\infer2[(S-ASG)]{\typenv,\typop \vdash \vv \asg \mpred{}(\vv): (\tierb,\tierb,\tierb) }
\hypo{\typenv(\z)=\tiera}
\infer1[(E-VAR)]{\typenv,\typop \vdash \z : (\tiera,\tierb,\tierb) }
\hypo{\typenv(\z)=\tiera}
\infer1[(E-VAR)]{\typenv,\typop \vdash \z : (\tiera,\tierb,\tierb) }
\hypo{\typenv(\uu)=\tierb}
\infer1[(E-VAR)]{\typenv,\typop \vdash \uu : (\tierb,\tierb,\tierb) }
\infer2[(E-OR)]{\typenv,\typop \vdash \X_2(\z \upharpoonright \uu) : (\tiera,\tierb,\tierb) }
\infer2[(S-ASG)]{\typenv,\typop \vdash \z \asg \X_2(\z \upharpoonright \uu) : (\tiera,\tierb,\tierb) }
\infer1[(S-SUB)]{\typenv,\typop \vdash \z \asg \X_2(\z \upharpoonright \uu) : (\tierb,\tierb,\tierb) }
\infer2[(S-SEQ)]{\pi_2'' \rightslice \typenv,\typop \vdash \vv \asg \mpred{}(\vv)\sep \z \asg \X_2(\z \upharpoonright \uu) : (\tierb,\tierb,\tierb) }
\end{prooftree}
}
$$
\hrulefill
\centering
\captionof{figure}{Typing derivation for the program ${\tt ce}$ of Figure~\ref{fig:ce}}\label{fig:typex}
                \end{minipage}};
\end{tikzpicture}
\end{figure}

\subsection{Intuitions}

We now give some brief intuition to the reader on the type discipline in the particular case where exactly two tiers, $\tiera$ and $\tierb$, are involved. The type system splits program variables, expressions, and statements between the two disjoint \emph{tiers}:
\begin{itemize}
\item $\tiera$ corresponds to a program component whose execution may result in an increase of the memory size and that cannot control the program flow (similar to safe inputs in~\cite{BelCoo92}).
\item $\tierb$ corresponds to a program component whose execution cannot result in a memory increase and that may control the program flow (similar to normal inputs in~\cite{BelCoo92}).
\end{itemize}

\noindent 
The type system of Figure~\ref{fig:TS} is composed of two sub-systems. The typing rules provided in Figure~\ref{sfig:tsprog} enforce that terms follow a standard simply-typed discipline. The typing rules of Figure~\ref{sfig:tsproc} will implement a standard non-interference type discipline \`a la Volpano et al.~\cite{VIS96} on the expression (and statement) tier, preventing data flows from tier $\tiera$ to tier $\tierb$. The transition between the two sub-type-systems is performed in the rule (P-DEC) of Figure~\ref{sfig:tsprog} that checks that the procedure body follows the tier-based type discipline once and for all in a procedure declaration.

In Figure~\ref{sfig:tsproc}, as tier $\tierb$ data cannot grow and are the only data driving the program flow, the number of distinct memory configurations on such data for a terminating procedure is polynomial in the size of the program input (\ie, number of symbols). Hence a typable and terminating procedure has a \emph{polynomial step count} in the sense of~\cite{C92}, \ie, on any input, the execution time of a procedure is bounded by a first-order polynomial in the size of their input and the maximal size of any answer returned by an oracle call. 

The innermost tier is used to implement a declassification mechanism on operators improving the type-system's expressive power: an operator may be typed differently depending on its calling context (the statement where it is applied). This is the reason why more than 2 tiers can be used in general.

 The outermost tier is used to ensure that oracles are only called on inputs of bounded size. This latter restriction on oracle calls enforces a semantic restriction, called \emph{finite lookahead revision}, introduced in~\cite{KS17,KS18} and requiring that, during each computation, the number of calls performed by the oracle on an input of increasing size is bounded by a constant.
 
 Let $\mpt$ be the class of second-order functionals computable by an oracle Turing machine with a polynomial step count and a finite lookahead revision.
\cite{KS18} shows that $\BFF=\lambda(\mpt)_2$. The type system of Figure~\ref{fig:TS} ensures that each terminating procedure of a well-typed program computes a function in $\mpt$.

\section{Characterizations of the class of Basic Feasible Functionals}

\subsection{Safe and terminating programs}
In this section, we show that typable ($\safe$) and terminating ($\SN$) programs capture exactly the class of basic feasible functionals ($\BFF$).

For a given set $\mathcal{S}$ of functionals having simple type over $\W$, let $\mathcal{S}_2$ be the restriction of  $\mathcal{S}$ to second-order functionals and let $\lambda(\mathcal{S})$ be the set of functions computed by closed simply-typed lambda terms using functions in $\mathcal{S}$ as constants. Formally, let $\lambda(\mathcal{S})$ be the set of functions denoted by the set of closed simply-typed lambda terms generated inductively as follows:
\begin{itemize}
\item for each type $\tau$, variables $x^\tau, y^\tau, \ldots$ are terms,
\item each functional $F \in \mathcal{S}$ of type $\tau$, $F ^\tau$ is a term,
\item for any term $t^{\tau'}$ and variable $x^{\tau}$, $\lambda x^\tau.t^{\tau'}$ is a term of type $\tau \to \tau'$,
\item for any terms $t^{\tau \to \tau'}$ and $s^\tau$, $t^{\tau \to \tau'}\! s^{\tau}$ is a term of type $\tau'$.
\end{itemize}
Each lambda term of type $\tau$ represents a function of type $\tau$ and terms are considered up to $\beta$ and $\eta$ equivalences. $\lambda(\mathcal{S})_2$ is called the \emph{second-order simply-typed lambda closure} of $\mathcal{S}$.

The second-order simply-typed lambda closure is monotone with respect to set inclusion.
\begin{lem}\label{lem:mon}
For any two classes of functionals $\mathcal{X},\mathcal{Y}$, if $\mathcal{X}\subseteq \mathcal{Y}$ then $\lambda(\mathcal{X})_2 \subseteq \lambda(\mathcal{Y})_2$.
\end{lem}
\noindent 
Procedures correspond to programs from~\cite{HKMP20}, we can redefine the class $\st$ from this paper in terms of procedures.
The semantics and type system restricted to procedures are equivalent, hence, given a program $\procedure_{\phi} = \mathtt{c}\ \ret \x$ belonging to the set of safe and terminating programs $\st$, we can write a program $$\prog = \boite{\Y, \overline{\y}}\letin{\procname(\X, \overline{\x})\{\mathtt{c}\ \ret \x\}}\callcc\ \procname(\Y, \overline{\y})$$ that computes the same functional as $\procedure_{\phi}$ and is typable and strongly normalizing provided the procedure is typable and terminating.

\begin{defi}[Safe and terminating procedures]
Given $\procedure \triangleq \procname(\overline{\X},\overline{\x})\{[\var \overline{\y}\sep\!]\ \cmd  \ \ret \x\}$, $\procedure$ is \emph{safe} if there exist a simple typing environment $\typenvs$, a safe operator typing environment and a triplet of tiers $(\sla,\sla_{in},\sla_{out})$ such that $\procedure$ is well-typed for these environments, {\ie}  $ \typenvs,\typop \vdash \cmd : (\sla,\sla_{in},\sla_{out})$ can be derived using the rules of Figure~\ref{fig:TS}.

Let $\sem{\procedure}$ be defined by $\sem{\procedure}(\overline{f},\overline{\w})=w \text{ iff } (\{\procedure\}, \store_\emptyset[\overline{\x} \leftarrow \overline{\w},\overline{\X} \leftarrow \overline{f}],\callcc\ \procname(\overline{\X},\overline{\x})) \toenv  \w$ (following Figure~\ref{fig:Com}).
$\sem{\procedure}$ is a second order partial functional in $(\W \to \W)^{\size{\overline{\X}}} \to \W^{\size{\overline{\x}}} \hookrightarrow \W$. 
If $\sem{\procedure}$ is a total function, then we say that the procedure $\procedure$ is terminating.

Let $\st$ be the set of safe and terminating procedures, and $\sem{\st}$ the set of functionals computed by those procedures.
\end{defi}

\noindent 
The characterization of $\BFF$ in terms of safe and terminating procedures discussed in the introduction can be stated as follows.
\begin{thmC}[\cite{HKMP20}]\label{thm:HKMP20}
$ \lambda(\sem{\st})_2 = \BFF.$
\end{thmC}

\noindent 
We now introduce an intermediate lemma stating that $\sem{\SN \cap \safe}$ is stable under second-order simply-typed lambda closure.
\begin{lem}\label{lem:lstable}
$\lambda(\sem{\SN \cap \safe})_2 =\sem{\SN \cap \safe}.$
\end{lem}
\begin{proof}
The inclusion $\sem{\SN \cap \safe} \subseteq \lambda(\sem{\SN \cap \safe})_2$ trivially holds by definition of the second-order simply-typed lambda closure. 

We show that $\lambda(\sem{\SN \cap \safe})_2  \subseteq  \sem{\SN \cap \safe}$ holds.
We first define a straightforward transformation $[\cdot]$ from lambda terms representing functions in $\lambda(\sem{\SN \cap \safe})_2$ to terms in $\SN \cap \safe$ as follows:
\begin{align*}
[x^\tau] &\triangleq \va_x\\
[\lambda x^\tau.t^{\tau'}] &\triangleq \lambda [x^\tau].[t^{\tau'}]\\
[t^{\tau \to \tau'}\ s^{\tau}] &\triangleq [t^{\tau \to \tau'}]\MVAt[s^{\tau}]\\
[F]&\triangleq \term, \text{with } F \in \sem{\SN \cap \safe},
\end{align*}
for $\term$ such that $\sem{\boite{\overline{\Y},\overline{\y}}{\letin{\overline{\procedure}} \term}} = F$. The existence of $\term$ is ensured by the definition of $\sem{\SN \cap \safe}$ but there is no uniqueness. For a given function $f \in \lambda(\sem{\SN \cap \safe})_2$, let $t_f$ be a lambda term representing $f$.
$[t_f]$ is the term obtained by applying the above transformation. Let $p_f$ be the corresponding list of procedure declarations obtained for each $F$ in the above transformation.
The function $f$ is of order $2$, hence its simple type must be of the shape $\underbrace{(\TW\to\TW)\to(\TW\to\TW)\cdots(\TW\to\TW)}_{k}\to\underbrace{\TW\to\TW\cdots\TW}_l\to\TW$.
The program $\prog= \boite{{\X_1,\ldots, \X_k},{\x_1, \ldots, \x_l}}{\letin{p_f} ([t_f]\MVAt {\X_1}\MVAt\cdots\MVAt\X_k)\MVAt {\x_1}\MVAt\cdots\MVAt\x_l}$ is such that $\sem{\prog} =\lambda {X_1, \ldots, X_k}.\lambda {x_1, \ldots, x_l}.(f\ {X_1}\ \cdots\ {X_k}\ {x_1}\ \cdots\ {x_l})  = f \in \lambda(\sem{\SN \cap \safe})_2$, as the second-order simply-typed lambda closure is closed under $\eta$-expansion. Moreover, the program $\prog$ is safe (as each procedure can be typed independently) and terminating (as it computes a total function). Hence $\sem{\prog} \in \sem{\SN \cap \safe}$.
\end{proof}

\noindent 
We are now ready to state a first characterization of $\BFF$ in terms of safe ($\safe$) and terminating ($\SN$) programs, showing that the external simply-typed lambda-closure of Theorem~\ref{thm:HKMP20} can be removed.
\begin{thm}\label{thm:BFF}
$ \sem{\SN \cap \safe}_2 = \BFF.$
\end{thm}
\begin{proof}
Any procedure $\procedure=\procname(\overline{\X},\overline{\x})\{\ldots \}$ can be transformed into a program:
$$\prog_\procedure \triangleq \boite{\overline{\Y},\overline{\y}}{\letin{\procname(\overline{\X},\overline{\x})\{\ldots \}}  \callcc\ \procname(\overline{\{\z \to \Y\MVAt\z\}},\overline{\y})},$$
that is, programs consisting of a single procedure call and a single procedure declaration and where oracle calls are restricted to type-1 program inputs $\overline{\Y}$ (basically the oracles).  Moreover, it is easy to check that if $\procedure \in \st$ then $\prog_\procedure \in \SN \cap \safe$, by construction.
Consequently, it holds that:
\begin{align*}
\BFF &= \lambda(\sem{\st})_2 &&(\text{By Theorem}~\ref{thm:HKMP20})\\
& \subseteq \lambda(\sem{\SN \cap \safe})_2 &&(\text{By Lemma}~\ref{lem:mon})\\
&= \sem{\SN\cap\safe} &&(\text{By Lemma}~\ref{lem:lstable})
\end{align*}
Now we show that $\sem{\SN \cap \safe} \subseteq  \lambda(\sem{\st})_2$. Any function $f \in \sem{\SN \cap \safe}$ is computed by a safe and terminating program $\prog$, \ie, $\sem{\prog}=f$. For a given simple type $\STW$, let $[\STW]^{*}$ be the type obtained as follows $[\TW]^{*} = \W$ and $[\STW \to \STW']^* = [\STW ]^* \to [\STW']^*$. We define by induction a straightforward transformation $[\cdot]^*$ that maps a program in normal form in $\SN \cap \safe$ to a function in $\lambda(\sem{\st})_2$:
\begin{align*}
[\va : \STW]^*  &\triangleq {x_{\va}}^{[\STW]^*}\\
[\lambda \va.\term : \STW \to \STW' ]^* &\triangleq \lambda [\va : \STW]^*.[\term : \STW']^* \\
[\term \MVAt \term' : \STW]^* &\triangleq [\term : \STW' \to \STW]^*\ [\term' : \STW']^*\\
[\callcc\ \procname(\overline{\clos},\overline{\term}) : \TW]^*&\triangleq  \sem{\prog_\procedure} [\overline{\clos} : \STW \to \STW]^* [\overline{\term} : \STW]^*\\
[\{ \x \to \term\} : \TW \to \TW]^* &\triangleq \lambda [\x : \TW]^*.[\term : \TW]^*\\
[\boite{\overline{\X} ,\overline{\x}}{ \ldots \kwfont{ in } \term}]^* &\triangleq \lambda[\overline{\X} : \overline{\TW \to \TW}]^*.\lambda[\overline{\x} : \overline{\TW}]^*.[\term]^*
\end{align*}
\noindent 
In the above definitions, we have sometimes omitted types for readability.
In the $\boite{}{}$ rule, the declarations do not get translated as procedures will be treated as function symbols from $\sem{\st}$ written as $\sem{\prog_{\procedure}}$.
Indeed, as $\prog_\procedure \in \SN \cap \safe$ it implies that the procedure $\procedure$ is terminating and safe, hence $\procedure\in\st$.
In the rule for application $\MVAt$, the type $\STW'$ can be guessed easily as we know that the term is simply-typed (the program is safe).
$[\prog]^* \in \lambda(\sem{\st})_2$ trivially holds as $[\prog]^*$ is a second-order functional consisting of a simply-typed lambda closure of functions $\sem{\prog_\procedure} \in \sem{\st}$.
Moreover, one can easily check that $f=\sem{\prog} =[\prog]^*$ and so the result follows by applying Theorem~\ref{thm:HKMP20}.
\end{proof}

\noindent 
We want to highlight that the characterization of Theorem~\ref{thm:BFF} is not just ``moving'' the simply-typed lambda-closure inside the programming language by adding a construct for lambda-abstraction. Indeed, the soundness of this result crucially depends on some choices on the language design that we have enforced: the restricted ability to compose oracles using closures, and the read-only mode of oracles inside a procedure call, implemented through continuations.

\subsection{Safe and terminating rank-\texorpdfstring{$r$}{r} programs}\label{ss:safe0}
The characterization of Theorem~\ref{thm:BFF} allows to use lambda abstractions in terms of the language.
However, we also show that this characterization is still valid in the absence of lambda-abstraction.

A safe program $\prog$ with respect to a typing derivation $\pi$ is a \emph{rank-$r$} program, if for any typing sub-derivation $\pi' \rightslice\ \pbla \lambda \va.{\term} : \STW$ of $\pi$, it holds that $\order(\STW) \leq r$. In other words, all lambda-abstractions are type-$k$ terms, for $k \leq r$. In particular, a rank-$(r+1)$ program, for $r\geq 1$, has variables that are at most type-$r$ variables. Rank-$0$ and rank-$1$ programs may have both type-0 and type-1 variables as these variables can still be captured by closures, procedure declarations, or boxes.

For a given set $S$ of well-typed programs, let $S_r$ be the subset of rank-$r$ programs in $S$, \ie, $S_r \triangleq \{ \prog \in S \ | \ \prog \text{ is a rank-$r$ program}\}$. For example, $\safe_r$ denotes the set of safe rank-$r$ programs.
It trivially holds that $\safe = \cup_{r \in \mathbb{N}}\safe_r$. The rank is clearly not uniquely determined for a given program. 
In particular, any rank-$r$ program is also a rank-$(r+1)$ program. Consequently, for any set $S$ of well-typed programs and any $i  \leq j$, it trivially holds that $S_i \subseteq S_j$.

\begin{exa}\label{ex:safe0}
Program {\tt ce} of Figure~\ref{fig:ce} is in $\safe_0$. Indeed, ${\tt ce} \in \safe$, \cf\ Example~\ref{ex:type}, and ${\tt ce}$ is a rank-$0$ program, as it does not use any lambda-abstraction.
\end{exa}

Now we revisit the syntax and semantics of safe rank-$0$ programs in $\safe_0$.
The programs are generated by the syntax of Figure~\ref{fig:synt}, where the terms are all type-$0$, denoted by $\term^0$, and redefined by:
$$
\text{Terms} \qquad \term^0 
   \rgl  \x \ | \ \X \MVAt {\term^0} \ | \ \callcc\ \procname(\overline{\clos},\overline{\term^0})
   $$
Moreover, there is no longer a need for call-by-name reduction in the big step operational semantics. 
As a consequence, the rules (TVar),  (OA), and (Call) of Figure~\ref{sfig:term} can be replaced  
by the following simplified rules:

\begin{figure}[!h]
$$
\begin{prooftree}
\hypo{\phantom{\term^0 \downarrow \x}}
\infer1[(TVar$^0$)]{(\sigma,\store,\x) \toenv \store(\x)}
\end{prooftree}
\quad 
\begin{prooftree}
\hypo{(\sigma,\store,\term_1^0 )\toenv \w}
\infer1[(OA$^0$)]{(\sigma,\store,\X \MVAt \term_1^0) \toenv \store(\X)(\w)}
\end{prooftree}
$$
\\[10pt]
$$
\begin{prooftree}
\hypo{(\sigma,\store,\overline{\term^0}) \toenv \overline{\w}}
\hypo{(\sigma,\store[\overline{\x} \leftarrow \overline{\w},\overline{\y} \leftarrow \overline{\emptyword}],\overline{\X} \mapsto \overline{\clos},\cmd ) \tost \store' }
\infer2[(Call$^0$)]{(\sigma \cup \{\procname(\overline{\X},\overline{\x})\{\var \overline{\y}\sep\!\ \cmd  \ \ret \z\} \},\store  \Imp \callcc\ \procname(\overline{\clos},\overline{\term^0})) \toenv \store'(\z)}
\end{prooftree}
$$
\end{figure}
\noindent 
We will show that any function in  $\BFF$ can be computed by a program in $\SN \cap \safe_0$.
For that purpose, we consider an alternative characterization of $\BFF$ that was introduced in~\cite{KS19} inspired by $PV^{\omega}$ from~\cite{CU93}.
$PV^{\omega}$ is defined as the set of simply-typed $\lambda$ terms defined from function symbols for each $\FP$ functions and a higher order recursor $\mathcal{R}$.
Instead of this recursor, we are going to use another bounded iterator $\mathcal{I}'$ that is shown in~\cite{KS19} to be equivalent to $\mathcal{R}$.
Let $\mathcal{I}'\colon(\pvW \to \pvW) \to \pvW \to \pvW \to \pvW \to \pvW$ be semantically defined by $\mathcal{I}'(F,a,b,c)=(\lambda x.F(lmin(x,a)))^{\size{c}}(b),$ where $lmin\in\pvW \to \pvW \to \pvW$ is defined by
$$lmin(a,b)=
\begin{cases}
a& \text{ if }\size{a}<\size{b}, \\
b& \text{ otherwise.}
\end{cases}
$$
Let $PV^{\omega}_2$ be the second-order restriction of $PV^{\omega}$.
\cite{CooKap89} defines $\BFF$ as the set of functions denoted by $PV^{\omega}_2$ terms and from~\cite{KS19} we have that $PV^{\omega}_2$ terms can be written as $\lambda \overline{X}.\lambda \overline{x}.\pvterm$, where $\pvterm$ is an order $0$ term of type $\pvW$ in \emph{normal form} that is defined inductively by:
$$\pvterm :: = x \ | \ X(\overline{\pvterm}) \ | \ F(\overline{\pvterm}) \ | \ \mathcal{I}'(\lambda z.\pvterm , \pvterm , \pvterm , \pvterm)$$
where $x$ and $z$ are order $0$ variable of type $\pvW$, $X$ is an order $1$ variable of type $\pvW^{n} \to \pvW$, $F \in \FP$ is an order $1$ function of type $\pvW^{{n}} \to \pvW$.
Using~\cite[Theorem 5.12]{CU93} extended to terms containing order $1$ free variables, we know that in the normal form of those terms $\pvterm$, the only $\lambda$ appearing are the $\lambda z.\pvterm$ used as the first argument of the recursor $\mathcal{I}'$.

\begin{prop}\label{thm:KS19}
$ \lambda(\{\mathcal{I}'\} \cup \FP)_2 = \BFF,$
where $\FP$ is the class of first-order polynomial time computable functions.
\end{prop}
\begin{proof}
From~\cite{CooKap89}, we know that the set of functions denoted by closed $PV^{\omega}_2$ terms corresponds to $\BFF$, that is $\BFF = \lambda(\{\mathcal{R}\}\cup\FP)_2$.
\cite{KS19} introduces an iterator $\mathcal{I}$ and the considered iterator $\mathcal{I}'$, and proves (\cite[Lemma~2.3]{KS19}) that $\lambda(\{\mathcal{R}\}\cup\FP)=\lambda(\{\mathcal{I}\}\cup\FP)$, then that $\lambda(\{\mathcal{I}\}\cup\FP)=\lambda(\{\mathcal{I}'\}\cup\FP)$ (\cite[Theorem~4.3]{KS19}).

We conclude that $\lambda(\{\mathcal{I}'\} \cup \FP)_2 = \BFF$.
\end{proof}
\noindent 
In order to prove completeness, let us now provide three examples of programs and show they belong to $\safe_0$.
The examples are a program computing the addition, a program simulating a polynomial time Turing machine, and a program computing the second order iterator $\mathcal{I}'$.

\lstset{language=,
morekeywords={declare, in, while, return, call, if, else, skip, @, var, box},
mathescape, frame=single, framesep=0pt, basicstyle=\upshape\ttfamily\small, lineskip=1pt, rulesepcolor=\color{gray!30},backgroundcolor=\color{gray!10},linewidth=\textwidth}

\begin{exa}[Addition]\label{ex:add}
  The following program computes the unary addition:

  \begin{lstlisting}
box [x,y] in
  declare add(u,v) {
      while (u != $\vide$) {
        u$ \asg \mpred{}$(u)$\sep$
        v$ \asg \msuc{1}$(v)
      }
      return v
  }
  in call add(x,y)
   \end{lstlisting}

This program is in $\safe_0$.
Indeed, it can be given the type $\TW \to \TW \to \TW$ for the safe operator typing environment $\typop$ defined by
\begin{align*}
\typop(\meq{})(\tierb) &\triangleq \{\tierb \to \tierb \to \tierb\} &\typop(\vide)(\tierb) &\triangleq \{\tierb\}\\
\typop(\mpred{})(\tierb) &\triangleq \{\tierb \to \tierb\} & \typop(\msuc{1})(\tierb) &\triangleq \{\tiera \to \tiera\},
\end{align*}
the procedure typing environment $\typproc$ defined by $\typproc(\mathtt{add}(\uu,\vv)\{\ldots\}) = \langle \typenv,(\tierb,\tierb,\tiera) \rangle$ with $\typenv(\uu) =\tierb$ and $\typenv(\vv)=\tiera$, and the simple typing environment $\typenvs$ such that $\typenvs(\uu)=\typenvs(\vv)=\TW$.
\end{exa}
Notice that any polynomial over unary words can be encoded by a typable program in $\safe_0 $ that follows a similar programming pattern and typing discipline.

\begin{exa}[Turing machines]\label{ex:tm}
Consider a one tape (deterministic) Turing Machine on a binary word.
Its tape can be encoded by variables $\x$ and $\y$, containing respectively the reversed part of the tape to the left of the head and the right part of the tape.
The head of the tape is pointing on the first symbol of $\y$, if any.
States are encoded by constant words, with a special $s^0$ word for the initial state
The current state is stored in the variable $\z$.

The machine running in time $\tt time$ on input $\tt input$ can be encoded by a procedure declaration, with a unique \verb:while: loop that will iterate $\size{\mathtt{time}}$ times the code that simulates one step transition of the machine.
This code consists in conditionals for testing the character under the head and the current state.
In the procedure ${\tt TM}$, of program $\prog_{{\tt TM}}$ below, 
we detail an example of a transition, namely
the case when the state is $s$ and read symbol is $0$. We assume that the transition function $\delta$ defines the next state as $s'$, that the symbol $1$ is written, and that the head moves to the right. In other words, $\delta(s, 0) = (s', 1, \text{right})$ is encoded.
Other cases follow the same pattern.

\begin{lstlisting}
box [t,d] in
  declare TM(time,input){
     var x,y,z$\sep$
     x $\asg$ $\vide \sep$
     y $\asg$ input $\sep$
     z $\asg$ $s^0 \sep$ //Initial state
     while (time $\meq{}$ $\vide$){
        if ($\head$(y) $\mequ{}$ 0){
           if (z $\mequ{}\ s$){ //transition (s, 0) -> (s', 1, right)
              $\z \asg s';$
              $\x \asg \msuc{1}(\x);$
              $\y \asg \mpred(\y) $
           }
           else {$\ldots$}$\sep$
           $\ldots$
        }
        $\ldots$
        time$ \asg \mpred$(time)
     }
     return y
  }
  in call TM(t,d)
  \end{lstlisting}

Procedure ${\tt TM}$ can be typed by a variable typing environment $\typenv$ and a safe operator typing environment $\typop$ such that:
\begin{align*}
 \forall e \in \{\x,\y,{\tt s},{\tt input}\},\ \typenv(e)&=\tiera , & \forall \op \in \{\meq{}, \msuc{1}, \mpred{}\},\ \tiera \to \tiera &\in  \Delta(\tierb)(\op),\\
 \typenv({\tt time})&=\tierb, & \forall \op \in \{\mequ{} ,\mpred{}, \head{}\},\ \tierb \to \tierb &\in \Delta(\tierb)(\op).
 \end{align*}
 Consequently, $\prog_{\tt TM} \in \safe_0$.
\end{exa}

As a consequence of Examples~\ref{ex:add} and~\ref{ex:tm}, any polynomial time TM can be simulated in $\safe_0$.

\begin{lem}\label{lem:ptmsafe0}
For any polynomial time Turing machine $M$, there exists a $\safe_0$ program that computes the same function as $M$.
\end{lem}
\begin{proof}
To prove this lemma, we use the procedure from Example~\ref{ex:tm}, encoding the execution of $|\mathtt{time}|$ steps of Turing machine $M$ on \verb:input:.
To give a correct bounding time, we assume that this machine runs in polynomial time, that is, there exists a polynomial $P$ such that on input $n$, the Turing machine halts in time bounded by $P(|n|)$.
As indicated in Example~\ref{ex:add}, we can write a safe procedure that computes $P$ in unary.
In other words, there is a procedure \texttt{poly} that takes \texttt{input} as argument and returns the unary representation of $P(|\mathtt{input}|)$.

The following program hence computes the same function as $M$.

\begin{lstlisting}
box [x] in
  declare poly(input){ ... }
  TM(time, input){ ... }
in call TM(call poly(x), x)
\end{lstlisting}

Both procedures have already been shown to be safe and the program has rank $0$.
\end{proof}

\begin{exa}[Second-order iterator]\label{ex:it}
The following program computes a variant (modulo padding) of the second-order iterator $\mathcal{I}'$ using the operator {\tt lmin} that computes the functional $lmin$ defined in Section~\ref{ss:safe0}.

\begin{lstlisting}
box [X,x$_1$,x$_2$,x$_3$] in
  declare It(Y,x,y,z){$\tikzmark{debut}$
    var u$\sep$
    u $\asg$ y $\sep$
    while (z $\meq{}\ \vide)${
      u$\asg $Y(lmin(u,x)$\upharpoonright$ x)$\sep$$\hspace{2cm}\tikzmark{tab}$
      z$ \asg \mpred($z$)$
    }
    return u
  }$\tikzmark{fin}$
  in call It(X,x$_1$,x$_2$,x$_3$)
  \end{lstlisting}
\AddNote{debut}{fin}{tab}{procedure $\mathtt{it}$}

This program can be shown to be in $\safe_0$ by setting a variable typing environment $\typenv$ such that $\typenv({\tt u}) =\typenv(\x)=\typenv(\y)=\tiera$ and $\typenv(\z)=\tierb$.
\end{exa}

We are now ready to characterize $\BFF$ in terms of safe and terminating rank-$0$ programs.

\begin{thm}\label{thm:BFF0}
$ \sem{\SN \cap \safe_0} =\BFF. $
\end{thm}
\begin{proof}
Lemma~\ref{lem:ptmsafe0} shows how polynomial time Turing machines can be simulated by a $\safe_0$ and $\SN$ program, hence for any $F\in\FP$, there exist procedures $\mathtt{p}^{F}$ and $\mathtt{tm}^{F}$ with name $\mathtt{Poly}^{F}$ and $\mathtt{TM}^{F}$ such that $\callcc\ \mathtt{TM}^{F}(\callcc\ \mathtt{Poly}^{F}(\x), \x)$ computes $F(\x)$.
Example~\ref{ex:it} exhibits how $\mathcal{I}'$ can be programmed in $\SN \cap \safe_0$ using a procedure $\mathtt{it}$ with name $\mathtt{It}$.
Moreover, for any procedure $\procedure$ thus defined, there exists a variable typing environment $\typenv$, a safe operator typing environment $\typop$, and a triplet of tiers $(\sla,\sla_{in},\sla_{out})$ such that the judgment $\pbl \stt(\procedure) :(\sla,\sla_{in},\sla_{out})$ can be derived and calls to procedures $\n(\procedure)$ terminate on all inputs as all the considered functions are total.
For a given functional denoted by $\lambda \overline{X}.\lambda \overline{x}.\pvterm \in PV^\omega_2$ in normal form, that is where all the $\lambda$ appearing in $\pvterm$ are the first argument of an $\mathcal{I}'$, let $\Fun(\pvterm)$ be a sequence of the procedures encoding the functions in $\{\mathcal{I}'\} \cup \FP$ that are applied in $\pvterm$. We use notations $\emptyset$ and $\sqcup$ to respectively denote the empty sequence and the concatenation of procedures without duplication.
Now we provide a transformation $[\cdot]^\dagger$ that maps a normal form of $PV^\omega_2$ to a program:
\small
\begin{align*}
\Fun(x_i) &\triangleq \emptyset & [x_i]^\dagger &\triangleq \x_i\\
\Fun(X_i) &\triangleq \emptyset & [X_i]^\dagger &\triangleq \X_i\\
\Fun(X(\pvterm_1,\ldots,\pvterm_k)) &\triangleq \bigsqcup \Fun(\pvterm_i) & [X(\pvterm_1,\ldots,\pvterm_k) ]^\dagger &\triangleq ( \ldots ([X]^\dagger \MVAt [\pvterm_1]^\dagger) \ldots)\MVAt [\pvterm_k]^\dagger \\
\Fun(F(\overline{\pvterm})) &\triangleq \Fun(\overline{\pvterm}) \sqcup \mathtt{tm}^F\sqcup \mathtt{p}^F & [F(\overline{\pvterm}) ]^\dagger &\triangleq \callcc\ \mathtt{TM}^{F}(\callcc\ \mathtt{Poly}^{F}([\overline{\pvterm}]^\dagger), [\overline{\pvterm}]^\dagger) \\
\Fun(\lambda z.\pvterm) &\triangleq \Fun(\pvterm) & [\lambda z.\pvterm]^\dagger &\triangleq \{ [z]^\dagger \to [\pvterm]^\dagger \}\\
\Fun(\lambda z.\pvterm_1,\pvterm_2, \pvterm_3, \pvterm_4) &\triangleq \bigsqcup \Fun(\pvterm_i)\sqcup \mathtt{it} & [\mathcal{I}'(\lambda z.\pvterm_1,\pvterm_2, \pvterm_3, \pvterm_4) ]^\dagger &\triangleq \callcc\ \mathtt{It}([\lambda z.\pvterm_1]^\dagger,[\pvterm_2]^\dagger,[\pvterm_3]^\dagger,[\pvterm_4]^\dagger) \\
\Fun({\pvterm_1}, \ldots, \pvterm_k) &\triangleq \bigsqcup \Fun(\pvterm_i) & [\pvterm_1, \ldots, \pvterm_k]^{\dagger} &\triangleq [\pvterm_1]^{\dagger}, \ldots, [\pvterm_k]^{\dagger}
\end{align*}
\normalsize
 $$[\lambda \overline{X}.\lambda \overline{x}.\pvterm]^\dagger \triangleq \boite{[\overline{X},\overline{x}]^\dagger}{\letin{\Fun(\pvterm)}{[\pvterm]^\dagger}}$$

For any $f$ denoted by normal form $\lambda \overline{X}.\lambda \overline{x}.\pvterm \in PV^\omega_2$, we have $\sem{[\lambda \overline{X}.\lambda \overline{x}.\pvterm]^\dagger}=f$. Moreover, $[\lambda \overline{X}.\lambda \overline{x}.\pvterm]^\dagger \in \SN \cap \safe_0$ as it is terminating and safe, and the transformation $[\cdot]^\dagger$ does not make use of any lambda-abstraction. Hence $f \in \sem{\SN \cap \safe_0}$ and, consequently, $\BFF \subseteq \sem{\SN \cap \safe_0}$.

Conversely, $\SN \cap \safe_0 \subseteq \SN \cap \safe$. Consequently,  $\sem{\SN \cap \safe_0} \subseteq \sem{\SN \cap \safe} = \BFF$, by Theorem~\ref{thm:BFF}.
\end{proof}

\noindent 
Hence the characterization of Theorem~\ref{thm:BFF} is just a conservative extension of Theorem~\ref{thm:BFF0}: lambda-abstractions, viewed as a construct of the programming language, allow for more expressive power in the programming discipline but do not capture more functions. As lambda-abstraction is fully removed from the programming language, this also shows that the simply-typed lambda closure of Theorem~\ref{thm:HKMP20} can be simulated through restricted oracle compositions in our programming language using closures and continuations. 

Moreover, the full hierarchy of safe and terminating rank-$r$ programs collapses.
\begin{cor}
$\forall r \in \mathbb{N}, \ \sem{\SN \cap \safe_r} = \BFF.$
\end{cor}

\subsection{Type inference}
Let the size $\size{\procedure}$ of the procedure $\procedure$ be the total number of symbols in $\procedure$.
Type inference for procedures is already known to be cubic in their size, as demonstrated in~\cite{HKMP20}. 

\begin{thmC}[\cite{HKMP20}]\label{thm:tiproc}
Given a procedure $\procedure$ and a safe operator typing environment $\Delta$, deciding if there exists a variable typing environment $\Gamma$ and a triplet of tiers $(\sla,\sla_{in},\sla_{out})$ such that $\pbl \stt(\procedure) :(\sla,\sla_{in},\sla_{out})$ holds can be done in time $\mathcal{O}(\size{\procedure}^3)$.
\end{thmC}
\noindent 
The proof of this result uses a reduction to 2-SAT with $\mathcal{O}(\size{\procedure}\times \sla^2)$ clauses, which can be solved in time linear in the number of clauses~\cite{EveItaSha76,AspPlaTar79}, for some tier $\sla \leq \size{\procedure}$.

Let the size $\size{\prog}$ of the program $\prog$ be the total number of symbols in $\prog$. Type inference is tractable for safe programs.

\begin{thm}\label{thm:tiprog}
Given a program $\prog$ and a safe operator typing environment $\Delta$,
\begin{itemize}
\item deciding whether $\prog \in \safe$ holds is a $\Ptime$-complete problem.
\item deciding whether $\prog \in \safe_0$ holds can be done in time $\mathcal{O}(\size{\prog}^3)$.
\end{itemize}
\end{thm}

\begin{proof}
Showing that $\prog \in \safe$ consists in finding a simple typing environment $\typenvs$ and a procedure typing environment $\typproc$ such that $\pbla  \prog : \overline{\TW \to \TW} \to \overline{\TW} \to \TW$ can be derived.
Now suppose that $\prog =\boite{\overline{\X},\overline{\x}}{\letin{\overline{\procedure}} \term}$. After $\size{\overline{\X},\overline{\x}}$ applications of the rule (P-BOX) of Figure~\ref{sfig:tsprog}, we end up into the following judgment
$\typenvs \uplus \{\overline{\X} : \overline{\TW \to \TW},\overline{\x} : \overline{\TW}\},\typproc,\typop \vdash  \letin{\overline{\procedure}} \term:  \TW.$
Deriving the above judgment consists of deriving:
$$\pi \rightslice \ \typenvs \uplus \{\overline{\X} : \overline{\TW \to \TW},\overline{\x} : \overline{\TW}\},\typproc,\typop \vdash  \term:  \TW.$$
and deriving $\size{\overline{\procedure}}$ judgments of the shape:
$$\pi_i \rightslice \typproc_1(\procedure_i),\Delta \vdash \stt(\procedure_i) : \typproc_2(\procedure_i)$$
with $1 \leq i \leq \size{\overline{\procedure}}$, $\overline{\procedure}=\procedure_1,\ldots,\procedure_{\size{\overline{\procedure}}}$, after $\size{\overline{\procedure}}$ applications of the rule (P-DEC) of Figure~\ref{sfig:tsprog}.
By Theorem~\ref{thm:tiproc}, we know that type inference can be done in time $\mathcal{O}(\size{\procedure_i}^3)$, for each derivation $\pi_i$. Hence, it follows that all the judgments $\pi_i$ can be inferred in time $\mathcal{O}(\size{\prog}^3)$.

It remains to study the type inference problem for the derivation $\pi$. By looking at Figure~\ref{sfig:tsprog}, the typing discipline for terms follows a standard simply-typed discipline augmented with constants (the procedure calls). It is well-known that type inference in the simply-typed lambda-calculus is a $\Ptime$-complete problem as any instance of the Circuit Value Problem (CVP) can be encoded in the former~\cite{M04}. For example, true and false can be encoded as the simple types $\lambda \x.\lambda \y.\x : \STW \to \STW' \to \STW$ and $\lambda \x.\lambda \y.\y :\STW \to \STW' \to \STW'$, respectively, and the Boolean disjunction can be encoded by $\lambda \x.\lambda \y.(\x\ (\lambda \x.\lambda \y.\x)\ \y) : ((\STW_1\to \STW_2 \to \STW_1) \to \STW_3 \to \STW_4) \to \STW_3 \to \STW_4$. The final type can be flatten to $\TW$ by performing an arbitrary number of applications. Hence on a term encoding a circuit in such a way, type inference corresponds to circuit evaluation.

For checking that a program is in $\safe_0$, as there are no lambda-abstraction, the only admissible types for variables are $\TW$ and $\TW \to \TW$ and the simple type inference of the above proof can be done in time linear in $\size{\prog}$ as it just consists in a linear syntactical check on programs. Hence the asymptotic complexity of checking that a program is in $\safe_0$ is the complexity of deriving the judgments $\pi_i$ for procedure bodies.
\end{proof}
\noindent 
Tractability of type inference is a nice property of the type system. Showing $\prog \in \SN$ is at least as hard as showing the termination of a first-order program, hence $\Pi_2^0$-hard in the arithmetical hierarchy. Therefore, the characterizations of Theorems~\ref{thm:HKMP20},~\ref{thm:BFF}, and~\ref{thm:BFF0} are unlikely to be decidable, let alone tractable. The purpose of the next Section will be to specify a tractable termination criterion preserving completeness of the characterizations of Theorem~\ref{thm:BFF} and Theorem~\ref{thm:BFF0}.

\section{A completeness-preserving termination criterion}

In this section, we show that the undecidable termination assumption ($\SN$) can be replaced with a criterion, called $\SCPS$, adapted from the Size-Change Termination (SCT) techniques of~\cite{LJBA01}, that is decidable in polynomial time and that preserves the completeness of the characterizations. 

\subsection{ Some considerations on program termination.}
We first show that studying the termination of a safe program can be reduced to the study of the procedure termination.
For that purpose, we need to define what is meant for a procedure to terminate by looking at the program semantics of Figure~\ref{sfig:term}. 

A procedure $\procedure =\procname(\overline{\X},\overline{\x})\{\var \overline{\y}\sep\!\ \cmd  \ \ret \z\}$ is \emph{terminating} (strongly normalizing) if for any $\overline{w} \in \overline{\W}$, any $\overline{f} \in \overline{\W \to \W}$, there exists $\vi \in \W$ such that the judgment $(\{\procedure \},\store[\overline{\Y} \leftarrow \overline{f}, \overline{\y} \leftarrow \overline{\w}]  \Imp \callcc\ \procname(\overline{\{\z \to \Y\MVAt\z\}},\overline{\y})) \toenv \vi$ can be derived.

\begin{lem}\label{lem:term}
For a given $\prog \in \safe$, if all procedures defined in $\prog$ terminate, then $\prog$ is terminating.
\end{lem}
\begin{proof}
If all procedures of a given program are terminating then any procedure call appearing in a term is terminating.
As the terms are simply-typed, they are known to be terminating, as a direct consequence of the strong normalization of the simply-typed lambda-calculus\ \cite{Tait67}. 
\end{proof}
\noindent 
Hence, ensuring the termination of each procedure of a given safe program is a sufficient condition for the program to terminate.
The converse trivially does not hold as, for example, a procedure with an infinite loop may be declared and not be called within a given safe program.

\subsection{Size-Change Termination}

SCT relies on the fact that if all infinite executions imply an infinite descent in a well-founded order, then no infinite execution exists.
To apply this fact for proving termination, \cite{LJBA01} defines Size-Change Graphs (SCGs) that exhibit decreases in the parameters of function calls and then studies the infinite paths in all possible infinite sequences of calls.
If all those infinite sequences have at least one strictly decreasing path, then the program must terminate for all inputs.
While SCT is $\PSpace$-complete, Ben-Amram and Lee~\cite{BAL07} develop a more effective technique, called SCP, that is in $\Ptime$.
The SCP technique is strong enough for our use case. In the literature, SCT and SCP are applied to pure functional languages.
As we shall enforce termination of procedures, we will follow the approach of~\cite{Avery06} adapting SCT to imperative programs.

First, we distinguish two kinds of operators that will enforce some (strict) decrease.
\begin{defi}
An $n$-ary operator $\op$ is \emph{(strictly) decreasing in} $i$, for $i \leq n$, if
$\forall \overline{\w} \in \overline{\W}$, $\overline{\w} \neq \overline{\emptyword}$, $\size{\sem{\op}(\overline{\w})} \leq \size{\w_i}$ ($\size{\sem{\op}(\overline{\w})} < \size{\w_i}$, respectively)
and $\sem{\op}(\overline{\emptyword}) = \emptyword$.
\end{defi}
\noindent 
For operators of arity greater than $2$, $i$ may not be unique but will be fixed for each operator in what follows. 

For simplicity, we will assume that assignments of the considered programs are \emph{flattened}, that is for any assignment $\x \asg \e$, either $\e= \y \in  \Var_0$, or $\e = \op(\overline{\x})$, with $\overline{\x} \in \overline{\Var_0}$, or $\e = \X(\y \upharpoonright \z)$, with $\y,\z \in \Var_0$ and $\X \in \Var_1$.
Notice that, by using extra type-$0$ variables, any program can be easily transformed into a program with flattened assignments, while preserving semantics and safety properties.

For each assignment of a procedure $\procedure$, we design a bipartite graph, called a SCG, whose nodes are type-$0$ variables in $(\local(\procedure) \cup \param(\procedure)) \cap \Var_0$ and arrows indicates decreases or stagnation from the old variable to the new. If a variable may increase, then the new variable will not have an in-arrow.

The bipartite graph is generated for any flattened assignment $\x \asg \e$ by:
\begin{itemize}
\item for each $\y$, $\y \neq \x$, we draw arrows from left $\y$ to right $\y$.
\item If $\e=\y$, we draw an arrow from left $\y$ to right $\x$.
\item If $\e=\op(\overline{\x})$, with $\op$ a:
\begin{itemize}
\item decreasing operator in $i$, we draw an arrow from $\x_i$ to $\x$.
\item strictly decreasing operator in $i$, we draw a ``down-arrow'' from $\x_i$ to $\x$.
\end{itemize}
\end{itemize}
\noindent 
In all other cases (neutral and non-decreasing operators, positive operators, oracle calls), no arrow is drawn. We will name this SCG graph G($\x \asg \e$).
Finally, for a set $V$ of variables, $G^V$ will denote the SCG obtained as a subgraph of $G$ restricted to the variables of $V$.

\begin{exa}\label{ex:scg}
Here are the SCGs associated to simple assignments of a procedure with three type-$0$ variables $\x,\y,\z$ using a strictly decreasing operator in $1$ ($\mathtt{pred}$), a decreasing operator in $2$ ($\mathtt{min}$), a positive operator (+1), and an oracle call.
$$
\begin{tabular}{|c|c|c|c|}
\hline
$\y \asg \mathtt{pred}(\x)$ &
$\y \asg \mathtt{min}(\x, \y)$ &
$\x \asg \x +1$ &
$\x \asg \X(\y \upharpoonright \z)$
\\
\hline
\begin{tikzpicture}
\matrix (m) [matrix of math nodes,row sep=0.5em,column sep=2em,minimum width=1em] 
{ \x & \x\\
  \y & \y\\
  \z & \z\\
  };
  \draw[->] (m-1-1) -- (m-1-2);
  \draw[->] (m-1-1) -- node {$\downarrow$} (m-2-2);
    \draw[->] (m-3-1) -- (m-3-2);
\end{tikzpicture}
&
\begin{tikzpicture}
\matrix (m) [matrix of math nodes,row sep=0.5em,column sep=2em,minimum width=1em] 
{ \x & \x\\
  \y & \y\\
  \z & \z\\
  };
  \draw[->] (m-1-1) -- (m-1-2);
  \draw[->] (m-2-1) -- (m-2-2);
  \draw[->] (m-3-1) -- (m-3-2);
\end{tikzpicture}
&
\begin{tikzpicture}
\matrix (m) [matrix of math nodes,row sep=0.5em,column sep=2em,minimum width=1em] 
{ \x & \x\\
  \y & \y\\
  \z & \z\\
  };
  \draw[->] (m-2-1) -- (m-2-2);
  \draw[->] (m-3-1) -- (m-3-2);
\end{tikzpicture}
&
\begin{tikzpicture}
\matrix (m) [matrix of math nodes,row sep=0.5em,column sep=2em,minimum width=1em] 
{ \x & \x\\
  \y & \y\\
  \z & \z\\
  };
  \draw[->] (m-2-1) -- (m-2-2);
  \draw[->] (m-3-1) -- (m-3-2);
\end{tikzpicture}\\
\hline
\end{tabular}
$$
\end{exa}

The language $\mathcal{L}(\cmd)$ of (potentially infinite) \emph{sequences of SCG} associated with the statement $\cmd$ is defined inductively as an $\infty$-regular expression.
\begin{align*}
\mathcal{L}(\x \asg \e) &\triangleq G(\x \asg \e) & \mathcal{L}(\ifa (\e) \{\cmd_1\}{\tt else}\{\cmd_2\}) &\triangleq \mathcal{L}(\cmd_1)+\mathcal{L}(\cmd_2)\\
\mathcal{L}(\cmd_1 \sep \cmd_2) &\triangleq \mathcal{L}(\cmd_1).\mathcal{L}(\cmd_2) &
\mathcal{L}(\while (\e) \{\cmd_1\}) &\triangleq \mathcal{L}(\cmd_1)^{\infty}
\end{align*}
where, following the standard terminology for automata~\cite{NP85}, $\mathcal{L}(\cmd)^{\infty}$ is defined by $\mathcal{L}(\cmd)^{\infty}\triangleq \mathcal{L}(\cmd)^{*}+\mathcal{L}(\cmd)^{\omega}$.
In the composition of SCGs, we are interested in paths that advance through the whole concatenated graph.
Such a path implies that the final value of the destination variable is of length at most equal to the initial value of the source variable.
If the path contains a down-arrow, then the length of the corresponding words decreases strictly.

\begin{defi}
Following the terminology of~\cite{BAL07}, a (potentially infinite) sequence of SCGs has a \emph{down-thread} if the associated concatenated graph contains a path spanning every SCG in the sequence and this path includes a down-arrow.
\end{defi}

\begin{exa}\label{ex:seq}
Consider the statement $$\cmd\triangleq \y \asg \mathtt{pred}(\x) \sep \y \asg \mathtt{min}(\x, \y) \sep  \x \asg  \x +1 \sep \x \asg \X(\y\! \upharpoonright\! \z),$$ whose SCGs are described in Example~\ref{ex:scg}.
The concatenated graph obtained from the (unique and finite) sequence of SCGs in $\mathcal{L}(\cmd)$ is provided below. It contains a down-thread (the path from $\x$ to $\y$).

$$
\begin{tabular}{|c|}
\hline
\begin{tikzpicture}
\matrix (m) [matrix of math nodes,row sep=0.5em,column sep=2em,minimum width=1em] 
{ \x & \x& \x& \x& \x\\
  \y & \y& \y& \y& \y\\
  \z & \z& \z& \z& \z \\
  };
  \draw[->] (m-1-1) -- (m-1-2);
  \draw[->] (m-1-2) -- (m-1-3);
  \draw[->] (m-1-1) -- node {$\downarrow$} (m-2-2);
  \draw[->] (m-2-2) -- (m-2-3);
  \draw[->] (m-2-3) -- (m-2-4);
  \draw[->] (m-2-4) -- (m-2-5);
  \draw[->] (m-3-1) -- (m-3-2);
  \draw[->] (m-3-2) -- (m-3-3);
  \draw[->] (m-3-3) -- (m-3-4);
  \draw[->] (m-3-4) -- (m-3-5);
\end{tikzpicture}
\\
\hline
\end{tabular}
$$

Note however that the concatenated graph associated to $\cmd\sep\cmd$ does not contain a down-thread.
\end{exa}
A (potentially infinite) sequence of SCGs is \emph{fan-in free} if the in-degree of nodes is at most 1. By construction, all the considered SCGs are fan-in free.

\subsection{Safety and Polynomial Size-Change}
Unfortunately, programs with down-threads can loop infinitely in the $\emptyword$ state.
To prevent this, we restrict the analysis to cases where while loops explicitly break out when the decreasing variable reaches $\emptyword$, that is procedures with while loops of the shape $\while(\x\ \meq{}\ \vide)\{\cmd\}$. 

\begin{defi}
For a given set $V$ of variables, we will say that $\cmd$ \emph{satisfies the simple graph property for} $V$ if for any while loop $\while(\x\ \meq{}\ \vide)\{\cmd'\}$ in $\cmd$ all sequences of SCGs $G_1^V G_2^V \ldots$ such that $G_1G_2 \ldots \in \mathcal{L}(\cmd')$ are fan-in free and contain a down-thread from $\x$ to $\x$.
A procedure is in $\SCPS$ if its statement satisfies the simple graph property for the set of variables in while guards.
A program is in $\SCPS$ if all its procedures are in $\SCPS$.
\end{defi}

\begin{exa}
The program ${\tt ce}$ of Figure~\ref{fig:ce} is in $\SCPS$. The language $\mathcal{L}(\stt(\procedure_{\tt KS}))$ corresponding to the body of procedure $\procedure_{\tt KS}$ is equal to $G_1.G_2.(G_3.G_4)^\infty$, where the SCGs $G_i$ are defined as follows:

$$
\begin{tabular}{|c|c|c|c|}
\hline
\scalebox{0.8}{$G_1$} &
\scalebox{0.8}{$G_2$} &
\scalebox{0.8}{$G_3$} &
\scalebox{0.8}{$G_4$}
\\
\scalebox{0.85}{$\uu \asg\X_1(\vide \upharpoonright \vide)$} &
\scalebox{0.85}{$\z \asg\vide$} &
\scalebox{0.85}{$\vv \asg \mpred{}(\vv)$} &
\scalebox{0.85}{$\z \asg \X_2(\z \upharpoonright \uu)$}
\\
\hline
\begin{tikzpicture}[every node/.style={scale=0.83}]
\matrix (m) [matrix of math nodes,row sep=0.5em,column sep=2em,minimum width=1em] 
{ \vv & \vv\\
  \uu & \uu\\
  \z & \z\\
  };
  \draw[->] (m-1-1) -- (m-1-2);
    \draw[->] (m-3-1) -- (m-3-2);
\end{tikzpicture}
&
\begin{tikzpicture}[every node/.style={scale=0.83}]
\matrix (m) [matrix of math nodes,row sep=0.5em,column sep=2em,minimum width=1em] 
{ \vv & \vv\\
  \uu & \uu\\
  \z & \z\\
  };
  \draw[->] (m-1-1) -- (m-1-2);
  \draw[->] (m-2-1) -- (m-2-2);
\end{tikzpicture}
&
\begin{tikzpicture}[every node/.style={scale=0.83}]
\matrix (m) [matrix of math nodes,row sep=0.5em,column sep=2em,minimum width=1em] 
{ \vv & \vv\\
  \uu & \uu\\
  \z & \z\\
  };
  \draw[->] (m-1-1) -- node {$\downarrow$} (m-1-2);
  \draw[->] (m-2-1) -- (m-2-2);
  \draw[->] (m-3-1) -- (m-3-2);
\end{tikzpicture}
&
\begin{tikzpicture}[every node/.style={scale=0.83}]
\matrix (m) [matrix of math nodes,row sep=0.5em,column sep=2em,minimum width=1em] 
{ \vv & \vv\\
  \uu & \uu\\
  \z & \z\\
  };
  \draw[->] (m-1-1) -- (m-1-2);
  \draw[->] (m-2-1) -- (m-2-2);
\end{tikzpicture}\\
\hline
\end{tabular}
$$
\\
First, the procedure body satisfies the syntactic restrictions on programs (flattened expressions and restricted while guards). 
Moreover, the procedure body satisfies the simple graph property for $\{\vv\}$ as there is always a down-thread on the path from $\vv$ to $\vv$ in $(G_3.G_4)^\infty$ and any corresponding sequence is fan-in free. Consequently, the program ${\tt ce}$ is in $\SCPS \cap \safe_0$, by Example~\ref{ex:safe0}.
\end{exa}

\begin{lem}\label{lem:subsct}
$\SCPS \cap \safe \subseteq  \SN$.
\end{lem}
\begin{proof}
All the procedures defined in a program of $\SCPS$ have down-threads for all executions and are terminating. Hence, by Lemma~\ref{lem:term}, the program is terminating.
\end{proof}
\noindent 
$\SCPS$ preserves completeness on safe programs for $\BFF$.
\begin{thm}\label{thm:sct}
$ \sem{\SCPS \cap \safe_0} = \sem{\SCPS \cap \safe} = \BFF.$
\end{thm}

\begin{proof}
Examples~\ref{ex:add},~\ref{ex:tm}, and~\ref{ex:it} 
illustrate that any function in $\{\mathcal{I}' \} \cup \FP$ can be computed by a program in $\safe_0$. The programs corresponding to these three examples are in $\SCPS$ as they mostly consist of procedures with loops of the shape $\while(\x\ \meq{}\ \vide)\{\cmd; \x \asg \mpred{}(\x)\}$, for some variable $\x$ ($\tt u$, $\tt time$, and $\tt z$, respectively) that is not assigned to in $\cmd$. Consequently, there is a down-thread from $\x$ to $\x$ and $\cmd$ satisfies the simple graph property on $\{\x\}$. We obtain that:
\begin{align*}
\BFF &= \lambda(\{\mathcal{I}'\} \cup \FP)_2 &&(\text{By Proposition}~\ref{thm:KS19})\\
& \subseteq \lambda(\sem{\SCPS\cap\safe_0})_2 &&(\text{By Examples}~\ref{ex:add},~\ref{ex:tm}, \text{ and}~\ref{ex:it})\\
& \subseteq \lambda(\sem{\SCPS\cap\safe})_2 &&(\text{By Lemma}~\ref{lem:mon})\\
&\subseteq \lambda(\sem{\SN\cap\safe})_2 &&(\text{By Lemmata}~\ref{lem:mon}\text{ and}~\ref{lem:subsct})\\
&= \sem{\SN\cap\safe} &&(\text{By Lemma}~\ref{lem:lstable})\\
&=\BFF &&(\text{By Theorem}~\ref{thm:BFF})
\end{align*}
Consequently, $\lambda(\sem{\SCPS\cap\safe_0})_2=\lambda(\sem{\SCPS\cap\safe})_2=\BFF$. Now it remains to notice that Lemma~\ref{lem:lstable} still holds when $\SCPS$ is substituted to $\SN$, \ie, $\lambda(\sem{\SCPS\cap\safe})_2 = \sem{\SCPS\cap\safe}$ and the result follows.
\end{proof}
\noindent 
While in general deciding if a program satisfies the size-change termination criterion is $\PSpace$-complete, $\SCPS$ can be checked in quadratic time and, consequently, we obtain the following results.

\begin{prop}\label{thm:scp}
Given a program $\prog$, deciding if $\prog \in \SCPS$ holds can be done in time $\mathcal{O}(\size{\prog}^2)$.
\end{prop}
\begin{proof}
Given a procedure $\procedure$, deciding if $\procedure \in \SCPS$ is a particular instance of SCP that \cite{BAL07} treats in time $\mathcal{O}(\size{\procedure}^2)$.
So deciding if the whole program is in $\SCPS$ can be done in $\mathcal{O}(\size{\prog}^2)$ by checking that each procedure is in $\SCPS$ (the size of each procedure being bounded by $\size{\prog}$).
\end{proof}

\begin{thm}\label{thm:scti}
Given a program $\prog$ and a safe operator typing environment,
\begin{itemize}
\item deciding whether $\prog \in \SCPS \cap \safe$ is a $\Ptime$-complete problem.
\item deciding whether  $\prog \in \SCPS \cap \safe_0$ can be done in time $\mathcal{O}(\size{\prog}^3)$.
\end{itemize}
\end{thm}
\begin{proof}
By Proposition~\ref{thm:scp} and the proof of Theorem~\ref{thm:tiprog}.
\end{proof}

\section{Conclusion and future work}\label{s:con}
We have presented a typing discipline and a termination criterion for a programming language that is sound and complete for the class of second-order polytime computable functionals, $\BFF$. This characterization has three main advantages: 
\begin{enumerate}
\item it is based on a natural higher-order programming language with imperative procedures; 
\item it is pure as it does not rely on an extra semantic requirements (such as taking the lambda closure); 
\item belonging to the set  $\SCPS \cap \safe$ can be decided in polynomial time. 
\end{enumerate} 
The benefits of tractability is that our method can be automated. However the expressive power of the captured programs is restricted. This drawback is the price to pay for tractability and we claim that the full SCT method, known to be $\PSpace$-complete, could be adapted in a more general way to our programming language in order to capture more programs at the price of a worse complexity. Moreover, any termination criterion based on the absence of infinite data flows with respect to some well-founded order could work and preserve completeness of our characterizations. Another issue of interest is to study whether the presented approach could be extended to characterize $\BFF$ in a purely functional language. 

\subsubsection*{Acknowledgments}Bruce M. Kapron's work was supported in part by NSERC RGPIN-2021-02481.
Emmanuel Hainry and Romain Péchoux's work was supported by the Inria associate team TC(Pro)$^3$.
\bibliographystyle{alphaurl}
\bibliography{bib}

\end{document}